\documentclass[a4paper,fleqn,usenatbib]{mnras}

\usepackage{newtxtext,newtxmath}

\usepackage[T1]{fontenc}
\usepackage{ae,aecompl}

\usepackage{subcaption}
\usepackage{graphicx, nicefrac}	
\usepackage{multicol}
\usepackage{amsmath}	
\usepackage{amssymb}	
\usepackage{nicefrac}
\usepackage{tablefootnote}
\usepackage{threeparttable}
\usepackage{nccmath}
\usepackage{xcolor}
\usepackage{upgreek}
\usepackage{dcolumn}
\usepackage{hyperref}

\title[31 GHz emissivity variations from $\rho$\,Oph]{Resolved observations at 31\,GHz of spinning dust emissivity variations in $\rho$\,Oph}

\author[C. Arce-Tord et al.]{Carla Arce-Tord,\!$^{1}$\thanks{E-mail: carce@das.uchile.cl} Matias Vidal,\!$^{2}$ Simon Casassus,\!$^{1}$ Miguel C\'{a}rcamo,\!$^{3,4}$ \newauthor Clive Dickinson,\!$^{4}$ Brandon S. Hensley,\!$^{5}$ Ricardo G\'{e}nova-Santos,\!$^{6,7}$ J. Richard Bond,\!$^{8}$ \newauthor Michael E. Jones,\!$^{9}$ Anthony C. S. Readhead,\!$^{10}$ Angela C. Taylor,\!$^{9}$ J. Anton Zensus \!$^{11}$\\\\
 $^{1}$Departamento de Astronom\'{\i}a, Universidad de Chile, Camino El Observatorio 1515, Las Condes, Santiago, Chile\\
 $^{2}$Universidad Aut\'onoma de Chile, Facultad de
 Ingenier\'ia, N\'ucleo de Astroqu\'imica \& Astrof\'isica, Av. Pedro
 de Valdivia 425,Providencia,\\ Santiago, Chile \\
 $^{3}$Departamento de Ingenier\'{i}a Inform\'{a}tica, Universidad de Santiago de Chile, Av. Ecuador, 3659, Santiago, Chile\\
 $^{4}$Jodrell Bank Centre for Astrophysics, Alan Turing Building, Dpt. of Physics and Astronomy, School of Natural Sciences, \\ The University of Manchester, Oxford Road, Manchester, M13 9PL, UK\\
$^{5}$Spitzer Fellow, Department of Astrophysical Sciences, Princeton University, Princeton, NJ 08544, USA\\
$^{6}$Instituto de Astrof\'{\i}sica de Canarias, E-38200 La Laguna, Tenerife, Canary Islands, Spain \\
$^{7}$Departamento de Astrof\'{\i}sica, Universidad de La Laguna (ULL), E-38206 La Laguna, Tenerife, Spain\\
 $^{8}$Canadian Institute for Theoretical Astrophysics, 60 St. George Street, University of Toronto, Toronto, ON, M5S 3H8, Canada\\
$^{9}$Sub-department of Astrophysics, University of Oxford, Denys Wilkinson Building, Keble Road, Oxford OX1 3RH, UK\\
$^{10}$Cahill Centre for Astronomy and Astrophysics, California Institute of Technology, Pasadena, CA 91125, USA\\
$^{11}$Max-Planck-Institut f\"ur Radioastronomie, Postfach 2024, 53010 Bonn, Germany\\
}

\date{Accepted XXX. Received YYY; in original form ZZZ}

\pubyear{2019}

\begin{document}

\label{firstpage}
\pagerange{\pageref{firstpage}--\pageref{lastpage}}
\maketitle

\label{firstpage}

\begin{abstract}
The $\rho$\,Oph molecular cloud is one of the best examples of spinning dust emission, first detected by the Cosmic Background Imager (CBI). Here we present 4.5\,arcmin observations with CBI\,2 that confirm 31\,GHz emission from $\rho$\,Oph\,W, the PDR exposed to B-type star HD\,147889, and highlight the absence of signal from S1, the brightest IR nebula in the complex. In order to quantify an association with dust-related emission mechanisms, we calculated correlations at different angular resolutions between the 31\,GHz map and proxies for the column density of IR emitters, dust radiance and optical depth templates. We found that the 31\,GHz emission correlates best with the PAH column density tracers, while the correlation with the dust radiance improves when considering emission that is more extended (from the shorter baselines), suggesting that the angular resolution of the observations affects the correlation results. A proxy for the spinning dust emissivity reveals large variations within the complex, with a dynamic range of 25 at 3$\sigma$ and a variation by a factor of at least 23, at 3$\sigma$, between the peak in $\rho$\,Oph\,W and the location of S1, which means that environmental factors are responsible for boosting spinning dust emissivities locally.

\end{abstract}

\begin{keywords}
radiation mechanism: general -- radio continuum: ISM -- ISM: clouds, ISM: individual objects: $\rho$\,Oph -- ISM: photodissociation region (PDR) -- ISM: dust
\end{keywords}

\section{Introduction}
Since 1996, experiments designed to measure CMB anisotropies have found a diffuse foreground in certain regions of our Galaxy in the range of 10-60\,GHz. For instance, the {\em Cosmic Background Explorer} (COBE) satellite measured a statistical correlation between the emission at 31.5\,GHz and 140\,$\upmu$m, at high Galactic latitudes and over large angular scales \citep{kog96}. The spectral index of this radio-IR correlated signal was explained by a superposition of dust and free-free emission. Later, \cite{lei97} detected diffuse emission at 14.5\,GHz, which was spatially correlated with IRAS 100 $\upmu$m. This emission was 60 times stronger than the free-free emission and well in excess of the predicted levels from synchrotron and Rayleigh-Jeans dust emission, thus confirming the existence of an Anomalous Microwave Emission (AME).

The AME is a dust-correlated signal observed in the 10-100\,GHz frequency range (see \citealt{dickinson18} for a review). Numerous detections have been made in our Galaxy by separating the AME excess from the other radio components, namely synchrotron, free-free, CMB and thermal dust emission. It has been distinctly measured by CMB experiments and telescopes at angular resolutions of $\sim$\,1$^{\circ}$; for instance, the WMAP \citep{bennett13} and \textit{Planck} \citep{planckdiff1, planckdiff2} satellites have provided detailed information on diffuse emission mechanisms in the Galaxy through full-sky maps. Based on such surveys, AME accounts for $\approx$\,30$\%$ of the Galactic diffuse emission at 30\,GHz \citep{planckdiff2}. AME has also been observed in several clouds and H{\sc ii} regions (e.g. \citealt{fin04, wat05, cass06, dav06}, \cite{dick07,dic09}, \citealt{ami09, dic09, planck, vid11, planckame14, genova15}), some of them being arcminute resolution observations of well known regions \citep{cas08, scaife10, castellanos_11, tibbs11, battistelli15}. There have been a few extra-galactic AME detections as well, such as in the galaxy NGC\,6946 \citep{murphy10, scaife_10_galaxy, hensley15, Murphy2018} and M\,31 \citep{planckm31, battistellim31}. Although, extra-galactic studies are still a challenge due to the diffuse nature of the emission. 

AME is thought to originate from dust grains rotating at GHz frequencies and emitting as electric dipoles. Long before its detection, \cite{erickson57} proposed that electric dipole emission from spinning dust grains could produce non-thermal radio emission. \citet{dra98} calculated that interstellar spinning grains can produce emission in the range from 10 to 100\,GHz, at levels that could account for the AME. Since then, additional detailed theoretical models have been constructed to calculate the spinning dust emission, taking into account all the known physical mechanisms that affect the rotation of the grains. The observed spectral energy distributions (SEDs) of a number of detections are a qualitative match to spinning dust models \cite[e.g.][]{planckame14}. Thus, the spinning dust (SD) hypothesis has been the most accepted to describe the nature of the AME. 

As SD emission depends on the environmental conditions, a quantitative comparison between AME observations and synthetic spectra can give us information on the local physical parameters. However, even when there is a dependence on the environmental parameters, variations over a wide range of environmental conditions (e.g. typical conditions for a dark cloud versus a photo-dissociation region) only produce emissivity changes smaller than an order of magnitude \citep[e.g.][]{dl98b, ali-haimoud09}. Although, greater changes in SD emissivity models are observed by varying parameters intrinsic to the grain population, as shown by \citet{hensley+17}, where they highlighted the importance of grain size, electric dipole and grain charge distributions. Observationally, \citet{tibbs_16} and \citet{Vidal2019} reported that they can only explain SD emissivity variations in different clouds by modifying the grain size distribution for the smallest grains. 

Within the SD model, the smallest grains are the largest contributors to the emission. These have sub-nanometer sizes and spin at GHz frequencies. Polycyclic Aromatic Hydrocarbons (PAHs) are natural candidates for the AME carriers, as they are a well established nanometric-sized dust population \citep{tielens08}, and correlations between AME and PAHs tracers have been observed \citep{scaife10, ysard, tibbs11, battistelli15}. However, a full-sky study by \cite{hensley16} found a lack of correlation between the \textit{Planck} AME map and a template of PAH emission constructed with 12\,$\upmu$m data. \citet{tibbs_13} found that in Perseus Very Small Grains (VSGs) emission at 24\,$\upmu$m is more correlated with AME than 5.8\,$\upmu$m and 8\,$\upmu$m PAH templates. Recently, \citet{Vidal2019} confirmed that 30\,GHz AME from the translucent cloud LDN\,1780 correlates better with a 70\,$\upmu$m map. Rotational nanosilicates have also been considered as the carriers of the AME and they can account for all the emission without the need of PAHs \citep{hoang16-silicate, hensley-draine17}. These results have raised questions about the PAHs hypothesis as the unique carriers of AME.

One of the key characteristics of AME, based on observations, is that it seems to be always connected with photo-dissociation regions (PDRs). PDRs are optimal environments for SD production \citep{dickinson18}: there is high density gas, presence of charged PAHs and a moderate interstellar radiation field (ISRF) that favours an abundance of small particles due to the destruction of large grains. For instance, \citet{pilleri+12} fitted ISO mid-IR spectra and showed that an increasing UV-radiation field in PDRs favours grains destruction. Also, \citet{habart03} used ISO observations and concluded that the emission in the $\rho$\,Oph\,W PDR is dominated by the photoelectric heating from PAHs and very small grains. The most significant AME detections have been traced back to PDRs \citep{cas08, planck, Vidal2019}. 

Among the most convincing observations on the SD hypothesis is the $\rho$\,Ophiuchi ($\rho$\,Oph) molecular cloud, an intermediate-mass star forming region and one of the brightest AME sources in the sky \citep{cas08, planck}. The region of $\rho$\,Oph exposed to UV radiation from HD\,147889, the earliest-type star in the complex, forms a filamentary PDR whose brightest regions in the IR include the $\rho$\,Oph\,W PDR \citep{liseau99, habart03}. $\rho$\,Oph\,W is the nearest example of a PDR, at a distance of 134.3\,$\pm$\,1.2\,pc in the Gould Belt \citep{gaiadr2}. In radio-frequencies, observations with the Cosmic Background Imager (CBI) by \cite{cas08} showed that the bright cm-wave continuum from $\rho$\,Oph, for a total \textit{WMAP}\,33\,GHz flux density of $\sim$20\,Jy, stems from $\rho$\,Oph\,W and is fitted by SD models. They found that the bulk of the emission does not originate from the most conspicuous IR source, S1. This motivated the study of variations of the SD emissivity per H-nucleus, $j_{\nu}$ = $I_{\nu}/N_{\mathrm{H}}$, which at the angular resolutions of their observations led to the tentative detection of $j_{\nu}$.

In this work, we report Cosmic Background Imager 2 (CBI\,2) observations of $\rho$\,Oph\,W, at a frequency of 31\,GHz and an angular resolution of 4.5\,arcmin. In relation to the CBI, this new data-set provides a finer angular resolution by a factor of 1.5 and better surface brightness sensitivity. Section\,\ref{sec:cbiobs} of this paper describes the CBI\,2 characteristics and observations, as well as the ancillary data. Sec.\,\ref{sec:morph} presents a qualitative analysis of the IR/radio features of the PDR. In Sec.\,\ref{sec:correl} we describe the results of the correlation analysis between the 31\,GHz data and the different IR and column density templates, and in Sec.\,\ref{sec:pahsize} we analyse the size variations of the PAHs in the PDR. In Sec.\,\ref{sec:emiss} we quantify the SD emissivity variations throughout $\rho$\,Oph and discuss the implications of our results, and Sec.\,\ref{sec:conc} concludes.

\section{The data}
\label{sec:cbiobs}
\subsection{Cosmic Background Imager 2 Observations} 
The observations at 31\,GHz were carried out with the Cosmic Background Imager 2 (CBI\,2), an interferometer composed of 13 Cassegrain antennae \citep{cbi2}. The CBI\,2 is an upgraded version of the Cosmic Background Imager \citep[CBI,][]{padin02}. In 2006, the antennae were upgraded from 0.9\,m (CBI) to 1.4\,m dishes (CBI\,2). As opposed to the CBI, CBI\,2 has an increased surface brightness sensitivity at small angular scales due to the incorporation of larger dishes that provide a better filling factor for the array. Technical specifications of the CBI and CBI\,2 are presented in Table\,\ref{tab:specific}. 

\begin{table}
	\centering
	\caption{Technical specifications for the CBI and CBI\,2.}
	\label{tab:specific}
	\begin{tabular}{lccr} 
		\hline
		  & CBI & CBI\,2\\
		\hline
		Years of operation & 1999-2006 & 2006-2008 \\
		Observing frequencies (GHz) & 26-36 & 26-36 \\
		Number of antennae & 13 & 13 \\
		Number of channels (1 GHz) & 10 & 10 \\
		Number of baselines & 78 & 78 \\
		Antenna size (m) & 0.9 & 1.4 \\
		Primary beam FWHM (arcmin) & 45 & 28.2 \\
		\hline
	\end{tabular}
\end{table}
 
The observations were spread over 27 days in March of 2008. We acquired data distributed in 6 pointings: 1 pointing centered at the $\rho$\,Oph\,W PDR, 3 pointings defined around the PDR, and 2 more pointings centered at stars S1 and SR3. Table\,\ref{tab:ophareas} shows the coordinates for each pointing, as well as their respective RMS noise. The data-set was reduced using the standard CBI pipeline (CBICAL) and software \citep{pearson03, readhead04, readhead04sci}, with Jupiter as the main flux calibrator. As the CBI\,2 is a co-mounted interferometer, the pointing error associated with each antenna has to be accounted by the combined pointing error of the baseline pairs of antennae. This yields a residual pointing error of $\sim$\,0.5\,arcmin \citep{cbi2}.

\begin{table}
	\centering
	\begin{threeparttable}
	\caption{Coordinates for the observation of the $\rho$\,Oph\,W PDR and the pointings defined around it. S1 and SR3 are the early-type stars of interest. The angular resolution of the observation is 4.5\,arcmin.}
	\label{tab:ophareas}
	\begin{tabular}{lccc} 
		\hline
		  & RA\tnote{a} & DEC\tnote{b} & RMS [mJy beam$^{-1}$] \\
		\hline
		$\rho$\,Oph\,W PDR & 16:25:57.0 & $-$24:20:48.0 & 6.7 \\
		$\rho$\,Oph\,1 & 16:26:06.0 & $-$24:37:51.0 & 9.3 \\
		$\rho$\,Oph\,2 & 16:26:00.0 &  $-$24:22:32.0 & 30.2 \\
		$\rho$\,Oph\,3 & 16:25:14.0 &   $-$24:11:39.0 & 15.0 \\
		$\rho$\,Oph\,S1 & 16:26:34.18 & $-$24:23:28.33 & 12.7 \\
		$\rho$\,Oph\,SR3 & 16:26:09.32 & $-$24:34:12.18 & 9.6 \\

		\hline
	\end{tabular}
	\begin{tablenotes}
    \item[a,b] J2000. Right ascension are in HMS and Declinations are in DMS.
    \end{tablenotes}
    \end{threeparttable}
\end{table}

\subsection[]{Image reconstruction} 
\label{sec:reconstruction}
\begin{figure}
\centering
	\includegraphics[width=0.48\textwidth,height=!]{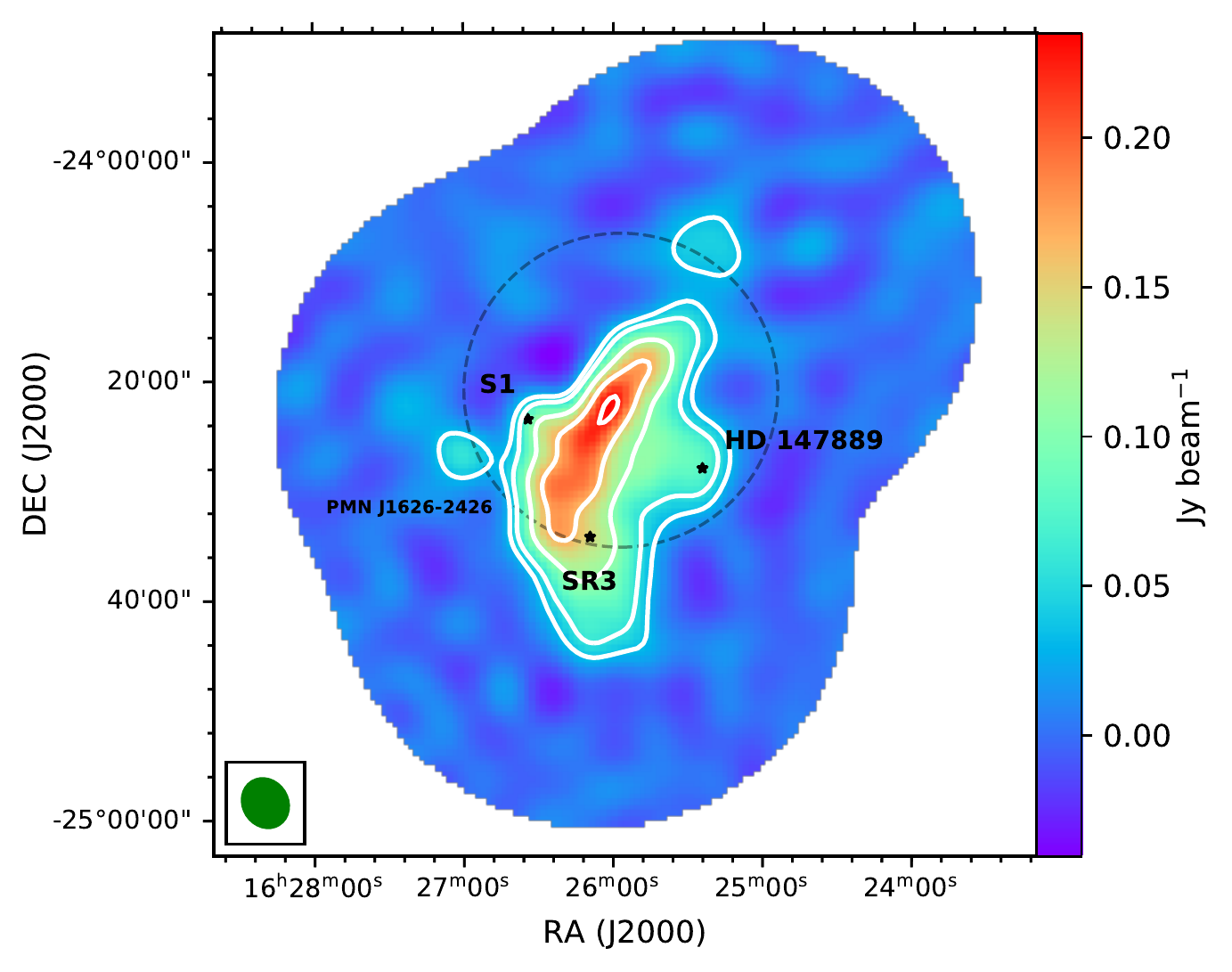}
    \caption{Mosaic of the 31\,GHz continuum measured by CBI\,2 in units of Jy\,beam$^{-1}$. This image was reconstructed using the CASA {\tt clean} task. The RMS noise is $\sigma \approx$0.01\,Jy\,beam$^{-1}$ and we show intensity contours for 3$\sigma$, 5$\sigma$, 10$\sigma$, 15$\sigma$ and 20$\sigma$. The star markers indicate the three early-type stars in the field: S1, SR3 and HD\,147889. We also detected radio galaxy PMN\,J1626-2426 (labeled in the figure). The CBI\,2 synthesized beam (4.6\,$\times$\,4.0\,arcmin) is shown as a green ellipse on the bottom left and the primary beam for the main pointing (28.2\,arcmin FWHM) is shown as a dashed circle.}
    \label{fig:cbi-points}
\end{figure}

We considered two approaches to reconstruct an image from the CBI\,2 data. One of them is the traditional {\tt clean} algorithm, in which we used a \textit{Briggs} robust 0 weighting scheme to construct a continuum mosaic. Figure\,\ref{fig:cbi-points} shows the reconstructed mosaic using the CASA\,5.1 {\tt clean} task \citep{casa}. This mosaic shows the location of the 3 early-type stars in the field - HD\,147889, S1 and SR3. It also presents the 31\,GHz intensity contours at 3$\sigma$, 5$\sigma$, 10$\sigma$, 15$\sigma$ and 20$\sigma$, for an RMS noise of $\sigma \approx$0.01\,Jy\,beam$^{-1}$. The dashed circle represents the 28.2\,arcmin primary beam of the central pointing ($\rho$\,Oph\,W PDR in Table\,\ref{tab:ophareas}), and the bottom left green ellipse shows the synthesized primary beam, with a size of 4.6\,$\times$\,4.0\,arcmin.

\begin{table}
	\centering
	\begin{threeparttable}
	\caption{Flux densities and spectral indexes for source PMN\,J1626-2426, detected at RA\,16:27:01.8, DEC\,-24:26:33.7 (J2000). The spectral index follows the convention S$_{\nu}$\,$\propto$\,$\nu^{\alpha}$ and it was calculated between the 31\,GHz flux from the {\tt clean} map and the catalogued fluxes. The reported fluxes correspond to different epochs, indicating variability in the spectrum.}
	\label{tab:pmn}

	\begin{tabular}{ccc} 
	\hline
	Observing Frequency [GHz] & Flux Density\,[mJy] & $\alpha_{(\nu/31\,\mathrm{GHz})}$ \\
    \hline
    1.4 & 57.2\,$\pm$\,1.8\,\tnote{a} & 0.01\,$\pm$\,0.03 \\
    4.8 & 132\,$\pm$\,13\,\tnote{b} & -0.43\,$\pm$\,0.07 \\
    8.4 & 77.9\,\tnote{c} &  -0.21 \\
    20 & 78\,$\pm$\,5\,\tnote{d} & -0.61\,$\pm$\,0.25 \\
    31 & 59.6\,$\pm$\,5.3\,\tnote{e} & - \\
		\hline
	\end{tabular}

	\begin{tablenotes}
    \item[a] Reported in \citet{pmn1.4}.
    \item[b] Reported in \citet{pmn4.8}.
    \item[c] Reported in \citet{pmn8.4}. Error not reported.
    \item[d] Reported in \citet{pmn20}. 
    \item[e] Reported in this work.
    
    \end{tablenotes}
    \end{threeparttable}
\end{table}

In a second approach we used an alternative image reconstruction strategy, based on  non-parametric image deconvolution, as in the maximum-entropy method \citep[following the algorithms described in ][]{cass06, cas08}. Here we used the {\tt gpu-uvmem} package \citep{carca}, implemented in GPU architectures. Thanks to entropy regularization and image positivity, the model image can super-resolve the {\tt clean} beam, in the sense that its effective angular resolution is $\sim$1/3 that of the natural-weighted beam (so even finer than uniform weights). To obtain a model sky image that fits the data we need to solve the usual deconvolution problem, i.e. obtain the model image $I^m$ that minimize a merit function $L$:  
\begin{ceqn}
\begin{align}
L = \chi^2 - \lambda \sum_i p_i\ln(p_i/M),
\end{align}
\end{ceqn}
where
\begin{ceqn}
\begin{align}
\chi^2 = \frac{1}{2} \sum_{k=0}^{N} \omega_k ||V_k^{\circ} - V_k^m||^2.
\end{align}
\end{ceqn}
Here, $V_k^{\circ}$ are the observed visibilities, each with weight $\omega_k$, and $V_k^m$ are the model visibilities calculated on the model images. The optimization is carried out by introducing the dimensionless free parameters $p_i = I_i/\sigma_D$, and $M$ is the minimum value for the free parameters. In this case, we set $M$\,=\,$10^{-3}$ and $\lambda$\,=\,2\,$\times$\,10$^{-2}$. The {\tt gpu-uvmem} algorithm also estimates a theoretical noise map as
\begin{ceqn}
\begin{align}
\sigma = \sqrt{\frac{1}{\sum_{p=1}^P \overline{A}^2_{p}/\hat{\sigma}^2}}\,,
\end{align}
\end{ceqn}
where $P$ is the number of pointings, $\overline{A}_{p} = \sum_{f=1}^F{A_{\nu_f}}/F$ is the mean of the primary beams over all frequencies $\{\nu_f\}_{f=1}^F$ available in pointing $p$, and $\hat{\sigma}$ is the noise on the dirty map using natural weights. Error propagation in the dirty map yields: 
\begin{ceqn}
	\begin{align}
	\hat{\sigma} = \sqrt{\frac{1}{\sum_{k} \omega_k}}\,.
	\end{align}
\end{ceqn}
In order to avoid an under-estimation of errors, we scaled the noise map so that its minimum value would match the RMS noise of the natural-weighted {\tt clean} map. In increasing the noise level we also included a correction for the number of correlated pixels in both maps, i.e. the noise in the {\tt gpu-uvem} model image is increased by $\sqrt{N_\mathrm{clean}/N_\mathrm{uvmem}} \sim 3$, where $N_\mathrm{clean}$ and $N_\mathrm{uvmem}$ are the number of pixels in each {\tt clean} beam, and that of {\tt gpu-uvmem} being $\sim$3 times smaller than the natural-weight beam.

In our analysis, we used both the {\tt clean} and {\tt gpu-uvmem} maps. The {\tt clean} map was used to measure the correlation of the 31\,GHz emission with the IR templates and the proxies for PAHs. We did not use the {\tt gpu-uvmem} map to perform the correlations as it has a variable angular resolution. Instead, as the {\tt gpu-uvmem} map recovers the morphology of the 31\,GHz emission in detail, we used it to construct an emissivity proxy to analyse the emissivity variations throughout the PDR.

In our reconstructed {\tt clean} mosaic we detected a point source located at RA\,16:27:01.8, DEC\,-24:26:33.7 (J2000) that corresponds to the radio galaxy PMN\,J1626-2426 (labeled in Figure\,\ref{fig:cbi-points}). This source is described in the PMN catalogue as a Flat Spectrum Radio Source, located at RA\,16:27:00.01, DEC\,-24:26:40.50 (J2000) \citep{pmnloc}. In Table\,\ref{tab:pmn} we report the flux and spectral index (S$_{\nu}$\,$\propto$\,$\nu^{\alpha}$) of this point source at 31\,GHz, along with catalog measurements at other frequencies. We measured the flux at 31\,GHz using aperture photometry with a circular aperture of radius 2.5\,arcmin. It is interesting to point out that this source has a \textit{Fermi} detection of 0.35\,$\pm$\,0.04\,pJy at 50\,GeV \citep{nolan12_fermi}, observed during 2008 and 2010. Its radio spectrum shows variability, especially considering the flux difference between the 1.4\,GHz data, observed during 1993, and the 4.8\,GHz data, observed during 1990 (see Table\,\ref{tab:pmn}). The 8.4\,GHz and 20\,GHz data were taken during 2003 and 2006, respectively.

\subsection{Ancillary Data} 
\label{sec:archival}
\begin{table}
	\centering
	\begin{threeparttable}
	\caption{Summary of the ancillary data.}
	\label{tab:ancillary}
	\begin{tabular}{lcl} 
		\hline
		 Instrument & Wavelength\,[$\upmu$m] & Description \\
		\hline
	\textit{Spitzer}-IRAC & 8 & Tracer for PAHs\tnote{a}. \\
    WISE & 3.4, 12 & Tracer for PAHs\tnote{a}. \\
	WISE & 22 & Tracer for VSGs. \\
	\textit{Spitzer}-MIPS & 24 & Tracer for VSGs. \\
	\textit{Herschel}-SPIRE & 250 & Tracer for big classical grains\,\tnote{b} \\
		\hline
	\end{tabular}

	\begin{tablenotes}
    \item[a] In $\rho$\,Oph\,W, this bands are dominated by PAHs. The 3.4$\upmu$m map is a better tracer for the smallest PAHs.
    \item[b] Grains with sizes of $\sim$ hundreds of nm in thermal equilibrium.
    \end{tablenotes}
    \end{threeparttable}
\end{table}

The Wide-Field Infrared Survey Explorer (WISE) has surveyed the entire sky in four near-IR and mid-IR band-passes \citep{wise}. Its near-IR survey at 3.4, 4.6 and 12\,$\upmu$m gives information on the emission from VSGs and PAHs in the interstellar medium. We worked with the 3.4\,$\upmu$m map, and the 12\,$\upmu$m processed map by \cite{meisner}, who smoothed the data-set to an angular resolution of 15\,arcsec and removed the compact sources and optical artifacts from the image. We also used their map at 22\,$\upmu$m for comparison purposes.

Dust grain tracers were also obtained from the InfraRed Array Camera ($\mathrm{IRAC}$) maps from the \textit{Spitzer} Space Telescope survey. $\mathrm{IRAC}$ is a mid-IR camera with four channels at 3.6, 4.5, 5.8, and 8\,$\upmu$m \citep{irac}. In this case, we used the 8\,$\upmu$m template. The original data-set is presented in units of MJy sr$^{-1}$ and at an angular resolution of $\sim$\,2 arcsec. We also used the 24\,$\upmu$m data-set from the \textit{Spitzer} Multiband Imaging Photometer ($\mathrm{MIPS}$) \citep{mips}. This photometer works at 24, 70 and 160\,$\upmu$m, producing maps with a native angular resolution between 6 and 40\,arcsec. 

In the particular case of $\rho$\,Oph\,W, both 8\,$\upmu$m and 12\,$\upmu$m bands are dominated by PAHs. Given the radiation field in HD\,147889 \citep[$\chi$\,$\sim$\,400,][]{habart03} and for typical dust models \citep{drli07, compiegne+11}, we expect both bands to be tracing primarily PAHs.

To study the thermal dust emission we used data from the \textit{Herschel} Gould Belt survey, which provides an extensive mapping of nearby molecular clouds, including detailed templates of $\rho$\,Oph \citep{herschel, herschel10}. Specifically, we used data from the Spectral Photometric Imaging Receiver ($\mathrm{SPIRE}$), a three-band camera and spectrometer that works at 250, 350 and 500\,$\upmu$m \citep{spire}. We also worked with the dust temperature data-set from the Photodetector Array Camera and Spectrometer ($\mathrm{SPIRE-PACS}$), that has a native angular resolution of 36.3\,arcsec \citep{ladjelate20}. A summary of our ancillary data is presented in Table\,\ref{tab:ancillary}.

\section{Qualitative radio/IR comparisons}
\label{sec:morph}
\begin{figure}
	\includegraphics[width=0.5\textwidth,height=!]{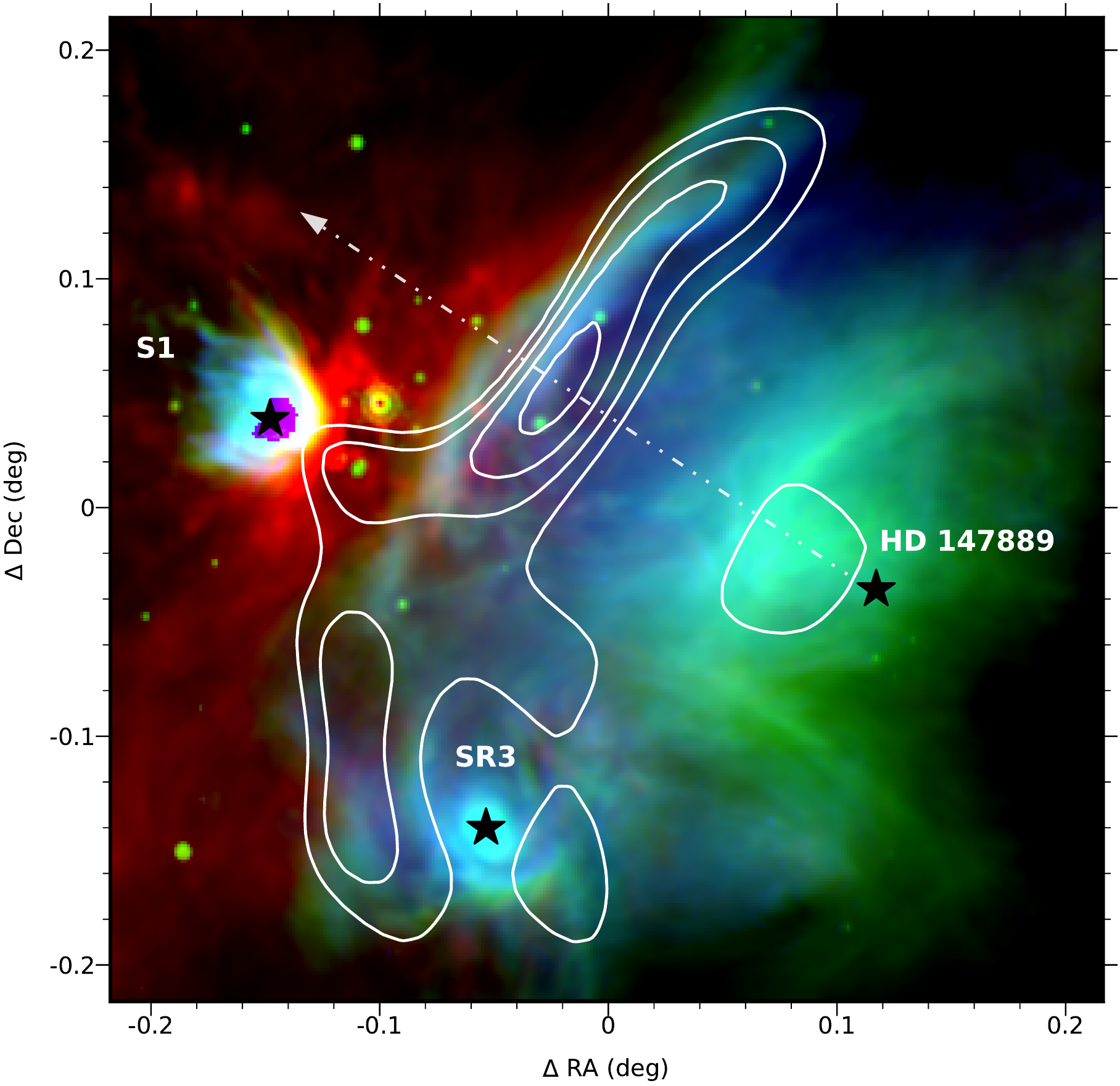}
    \caption{Overview of $\rho$\,Oph. Red: SPIRE\,250\,$\upmu$m, green: MIPS\,24\,$\upmu$m, blue: IRAC\,8\,$\upmu$m (PDR). The center of the image is at RA\,16:25:55.20, DEC\,-24:25:48.0 (J2000). The markers indicate the locations of the 3 early-type stars in the field. Contours follow a {\tt gpu-uvmem} model of the 31\,GHz continuum measured by the CBI\,2, at 30$\%$, 50$\%$, 70$\%$ and 90$\%$ of the emission peak at 0.51 Jy\,beam$^{-1}$. The white dashed arrow indicates the direction of the cut used to construct the emission profiles of the templates.}
    \label{fig:globalrgb}
\end{figure}

In Figure\,\ref{fig:globalrgb} we compare the CBI\,2 31\,GHz emission reconstructed with the {\tt gpu-uvmem} algorithm (Sect.\,\ref{sec:reconstruction}) shown in contours, with the $\mathrm{IRAC}$\,8\,$\upmu$m in blue colour stretch, $\mathrm{MIPS}$\,24\,$\upmu$m in green, and $\mathrm{SPIRE}$\,250\,$\upmu$m in red. The 8 and 24\,$\upmu$m templates trace the emission of small hot grains (sub-nm and nm-sizes), while the 250\,$\upmu$m template traces thermal dust emission, mainly from big classical grains (size of hundreds of nm). In particular, the $\mathrm{IRAC}$ 8\,$\upmu$m template is a main tracer of PAHs in this region. 

\begin{figure}
\begin{subfigure}{\columnwidth}
\includegraphics[width=\linewidth]{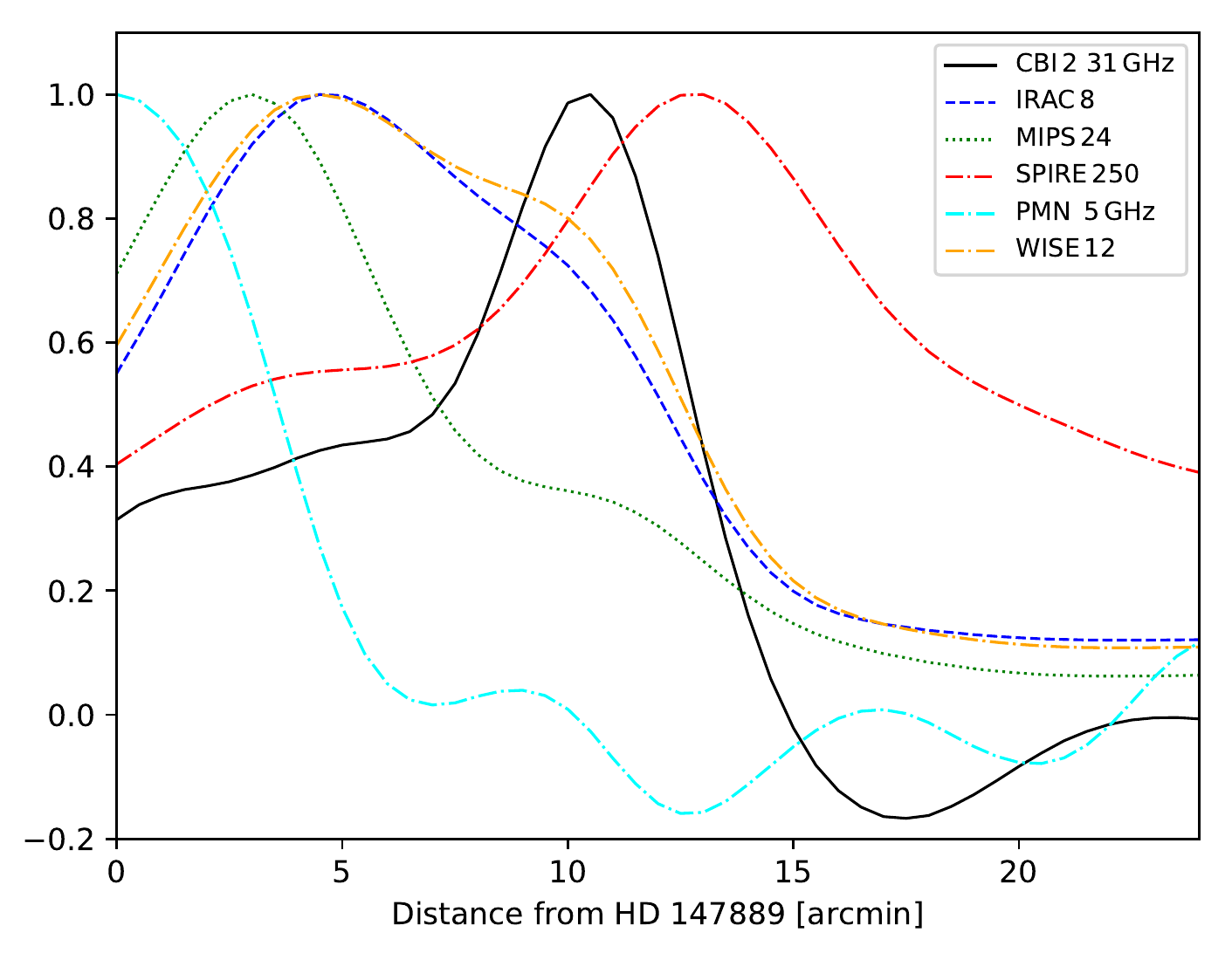}
\caption{} \label{fig:slit_ir}
\end{subfigure}
\medskip
\begin{subfigure}{\columnwidth}
\includegraphics[width=\linewidth]{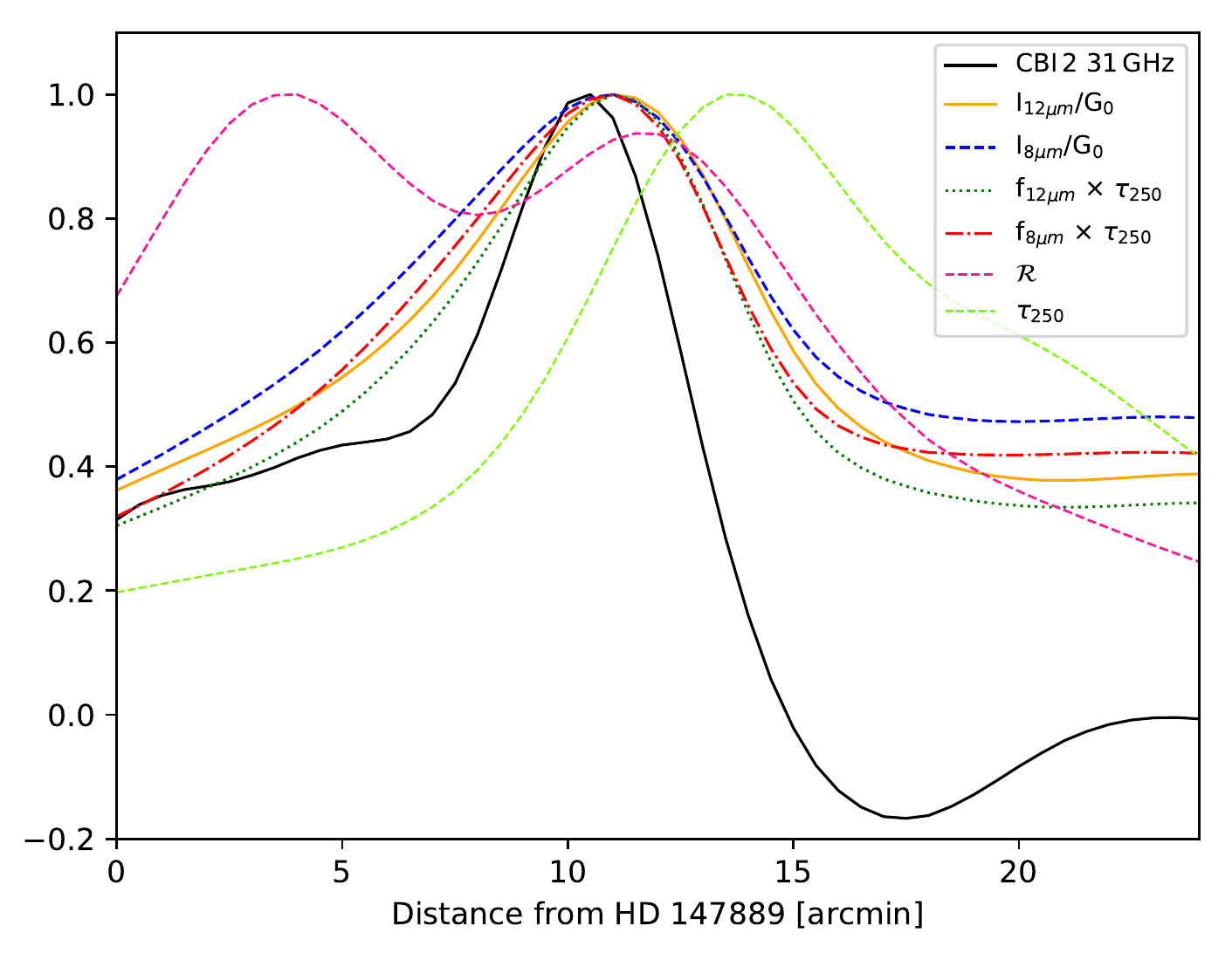}
\caption{} \label{fig:slit_proxies}
\end{subfigure}
\caption{Normalized emission profiles vs. distance to ionizing star HD\,147889. The profile cut, as shown by the dashed arrow in Fig.\,\ref{fig:globalrgb}, was extracted starting from star HD\,147889 and crossing the 31\,GHz emission peak.} \label{f:slits}
\end{figure}

The black markers in Fig.\,\ref{fig:globalrgb} show the position of the three early-type stars in $\rho$\,Oph. S1 is a binary system composed of a B4V star and a K-type companion, with an effective temperature of $T_{\mathrm{eff}}\approx$\,15800\,K. SR3, also known as Elia\,2-16, is a B6V star with an effective temperature of $T_{\mathrm{eff}}\approx$\,14000\,K. HD\,147889, the main ionizing star of $\rho$\,Oph, is catalogued as a single B2III/B2IV star \citep[C-ionizing,][]{houk88}, with  $T_{\mathrm{eff}}\approx$\,23000\,K \citep{bondar12}. This star is heating and dissociating gas layers creating an H{\sc ii} region of $\sim$\,6\,arcmin towards the PDR traced by the free-free continuum in the PMN\,(Parkes-MIT-NRAO)\,4.85\,GHz map, and by the H$\alpha$ emission in SHASSA \citep{cas08}. In Figure\,\ref{fig:slit_ir} we show normalized emission profiles for the IR templates and the 31\,GHz data, for a cut starting from HD\,147889 and passing through the 31\,GHz emission peak, as shown by the dashed arrow in Fig.\,\ref{fig:globalrgb}. We also plot the emission profile from the PMN\,4.85\,GHz map as a tracer of the free-free emission throughout the region. We find that, indeed, most of the 4.85\,GHz emission comes from the vicinity of HD\,147889. The 24\,$\upmu$m map in Fig.\,\ref{fig:globalrgb} also shows thermal
emission from hot grains near HD\,147889. Towards the PDR (at $\sim$\,6\,arcmin in Figure \ref{f:slits}) and through the peak of the 31\,GHz emission, the 4.85\,GHz emission decreases, reaching the noise level of the original map. This lack of free-free emission at the location of the PDR can also be seen in the 1.4\,GHz and H$\alpha$ maps of $\rho$\,Oph presented in \citet{planck}.  

An interesting aspect of Fig.\,\ref{fig:globalrgb} is that the 31\,GHz contours fall in the transition between small hot grains and bigger colder grains (reflected in the layered structure of the IR tracers, as also seen in Fig.\,\ref{fig:slit_ir}). This transition occurs at the PDR, where neutral Hydrogen becomes molecular. Deeper into the molecular core, away from HD\,147889 in Fig.\,\ref{fig:globalrgb}, the UV radiation field is attenuated, and the emergent emission progressively shifts towards 250\,$\upmu$m. 
 
Figure\,\ref{fig:slit_ir} shows that $\mathrm{IRAC}$\,8\,$\upmu$m and $\mathrm{WISE}$\,12\,$\upmu$m have a wider radial profile than that at 31\,GHz, ranging from their peak at $\sim$5\,arcmin distance from HD\,147789 up to $\sim$15\,arcmin. Note that around 5\,arcmin there is a slight hump on the 31\,GHz profile that matches the peak of $\mathrm{IRAC}$\,8\,$\upmu$m. Also, at $\sim$10\,arcmin, the $\mathrm{WISE}$\,12\,$\upmu$m profile shows a hump at the position of the 31\,GHz peak. This correlations might imply a relation between the rotational excitement of dust grains that originate the SD emission and the vibrational states of the small grains seen in the 8\,$\upmu$m and 12\,$\upmu$m profiles.

We note that the most conspicuous feature in the IR-map in Fig.\,\ref{fig:globalrgb}, the circumstellar nebula around S1, shows a very faint 31\,GHz counterpart. This is particularly interesting as \textit{Spitzer} IRS spectroscopy shows very bright PAHs bands in the nebulae around S1 and SR3, as well as in the $\rho$\,Oph\,W PDR \citep{habart03, cas08}. This means that PAHs are not depleted around S1 and SR3, thus making the faint 31\,GHz emission intriguing. Motivated by this, we performed a correlation analysis in order to quantify the best tracer of the 31\,GHz emission throughout the region.

\section{Correlation Analysis} 
\label{sec:correl}
We study the morphological correlation between the 31\,GHz data and different IR maps that trace PAHs and VSGs. In order to avoid biases due to image reconstruction, we performed the correlations in the visibility ($uv$) plane. To transform our data-set to the $uv$-plane we used the $\mathrm{MOCKCBI}$ software from the CBI software tools. This program calculates the visibilities by Fourier-transforming a supplied map of the sky emission, in this case, various IR templates. To do this, all templates were reprojected to a 1024\,$\times$\,1024 grid. In case that any of the original templates were of a smaller size, a noise gradient, calculated with the RMS noise limits of each map, was added to the borders. This is an important step to consider, otherwise, the resulting visibility maps would show an abrupt step towards its borders, producing artefacts in the mock visibilities. To visualize and compare the trend of our results we also calculated the correlations in the plane of the sky. We smoothed all the templates to 4.5\,arcmin in order to fit CBI\,2's angular resolution. All templates were re-gridded to $\sim$\,3\,pixels per beam to avoid correlated pixels.

\subsection{PAH column density proxies}
For the correlation analysis, we constructed proxies for the column density of PAHs. The mid-IR dust emission, due to PAHs, depends on the column density of the emitters and on the intensity of the local UV radiation, which can be quantified in units of the ISRF in the solar neighbourhood as the dimensionless parameter G$_{0}$ \citep{sellgren85, drli07}. If the observed 31\,GHz emission corresponds to SD emission, a stronger correlation is expected with the mid-IR emission when divided by G$_{0}$, as this will trace the column density of PAHs. 

The radiation field intensity was estimated using the equation given by \citet{ysard}:
\begin{ceqn}
\begin{align}
G_{0}\,=\,\left(T_{\mathrm{dust}}/17.5 K\right)\,^{\beta_{\mathrm{dust}} + \ 4}.
\label{ysard-g0}
\end{align}
\end{ceqn}
This method derives from radiative equilibrium with a single grain size of 0.1\,$\upmu$m and an emissivity index\footnote{This is the spectral index of the grain opacity.} $\beta_{ \rm dust}$\,=\,2, which is constant across the field. As $\beta_{ \rm dust}$ may vary throughout the PDR, a constant $\beta$ might not be a good approximation for G$_{0}$. We estimated the impact of variable $\beta_{\rm dust}$ in Eq.\,\ref{ysard-g0}. We created a map of $\beta_{\rm dust}$ and T$_{\rm dust}$ by fitting a Modified Black Body to the far-IR data. For this, we used Herschel data at 100, 160, 250, 350 and 500\,$\upmu$m. This fit results in $\beta_{ \rm dust}$, $\tau$ and T$_{ \rm dust}$ maps that would let us identify variations in the PDR at 4.5\,arcmin. The resulting G$_{0}$ map, calculated with the variable $\beta_{ \rm dust}$ and T$_{ \rm dust}$ maps from the fit, is morphologically equivalent to the G$_{0}$ estimation using a constant $\beta_{ \rm dust}$ and Herschel's T$_{ \rm dust}$ map. The Pearson correlation coefficient is $r$\,=\,0.95\,$\pm$\,0.02 between both G$_0$ estimations. In this work, we are interested in comparing the morphology of the cm-wavelength (31\,GHz) signal with that of the average radiation field along the line of sight. A 3D radiative transfer model that accounts for the IR spectral variations would provide an accurate estimation of the 3D UV radiation field (G$_0$(${\vec{x}}$)) and of the dust abundances \citep[e.g., as in][]{galliano+18}. Such modeling  is beyond the scope of this work.

In stochastic heating, we expect the intensity of the IR bands due to PAHs to be approximately proportional to both the UV field intensity and the PAH column, so that \citep[as in][]{cas08}:
\begin{ceqn}
\begin{align}
\frac{I_{\nu\,\mathrm{PAH}}}{G_0} \propto N_{\mathrm{PAH}} .
\label{colcass}
\end{align}
\end{ceqn}
As templates for the PAHs emission, $I_{\nu\,\mathrm{PAH}}$, we used the $\mathrm{WISE}$\,12\,$\upmu$m and $\mathrm{IRAC}$\,8\,$\upmu$m intensity maps.

\begin{figure*} 
\begin{subfigure}{0.33\textwidth}
\includegraphics[width=\linewidth]{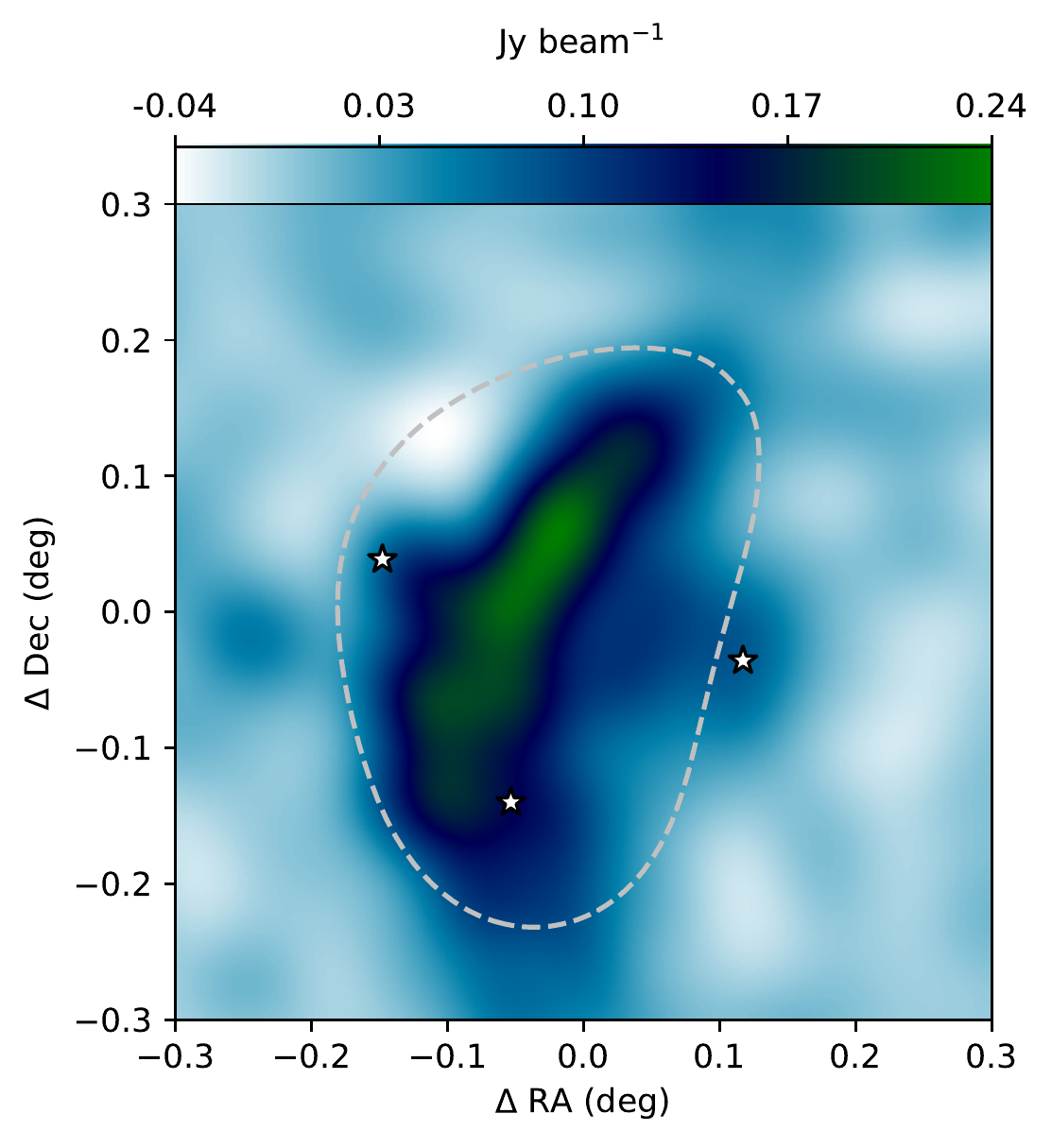}
\caption{CBI\,2 31 GHz} \label{fig:a}
\end{subfigure}\hspace*{0.5em}
\begin{subfigure}{0.33\textwidth}
\includegraphics[width=\linewidth]{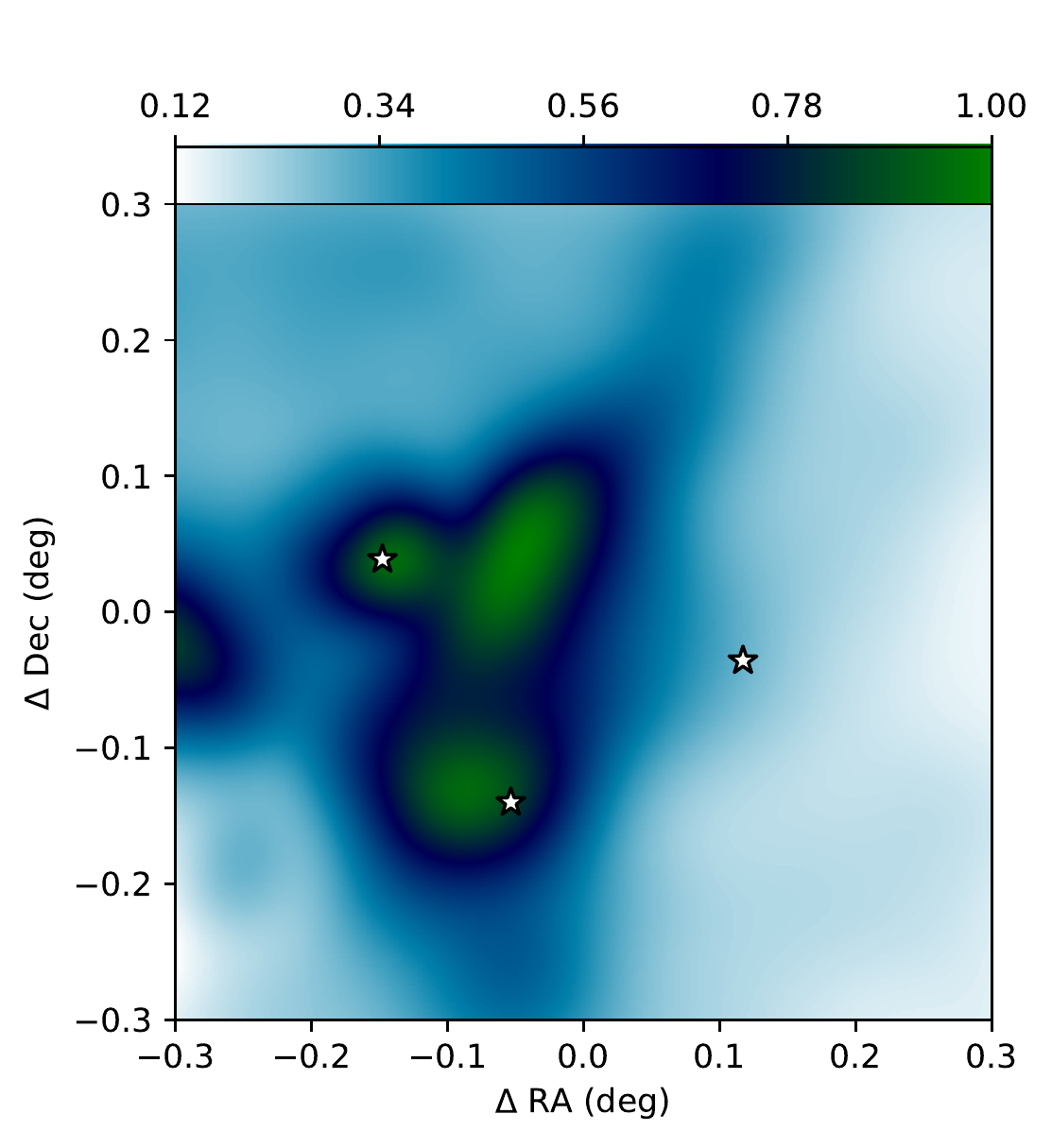}
\caption{$f_{\mathrm{\,12\,\upmu m}} \, \cdot \, \tau_{250}$} \label{fig:b}
\end{subfigure}\hspace*{0.5em}
\begin{subfigure}{0.33\textwidth}
\includegraphics[width=\linewidth]{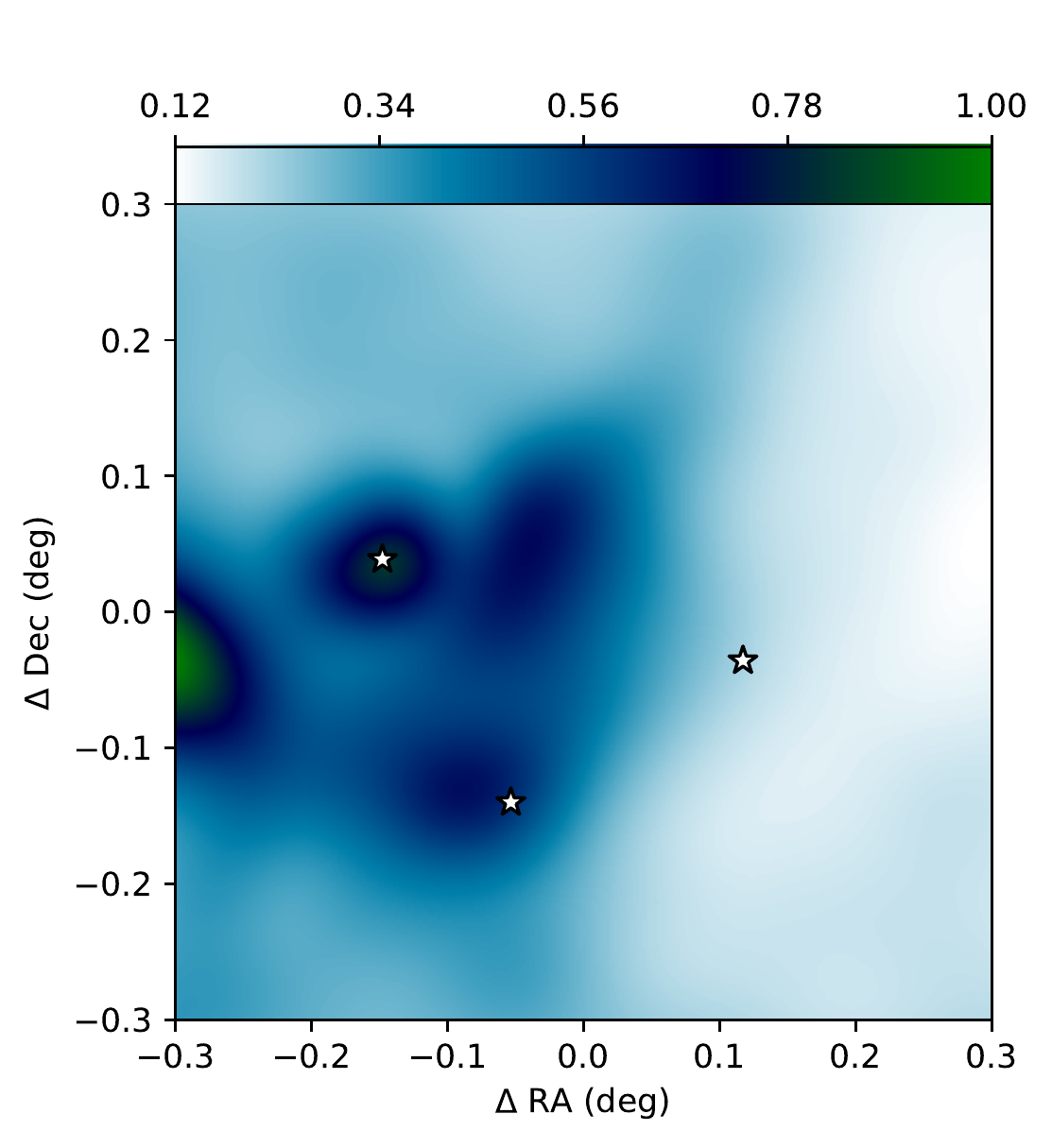}
\caption{$f_{\mathrm{8\,\upmu m}} \, \cdot \, \tau_{250}$} \label{fig:c}
\end{subfigure}
\medskip
\begin{subfigure}{0.33\textwidth}
\includegraphics[width=\linewidth]{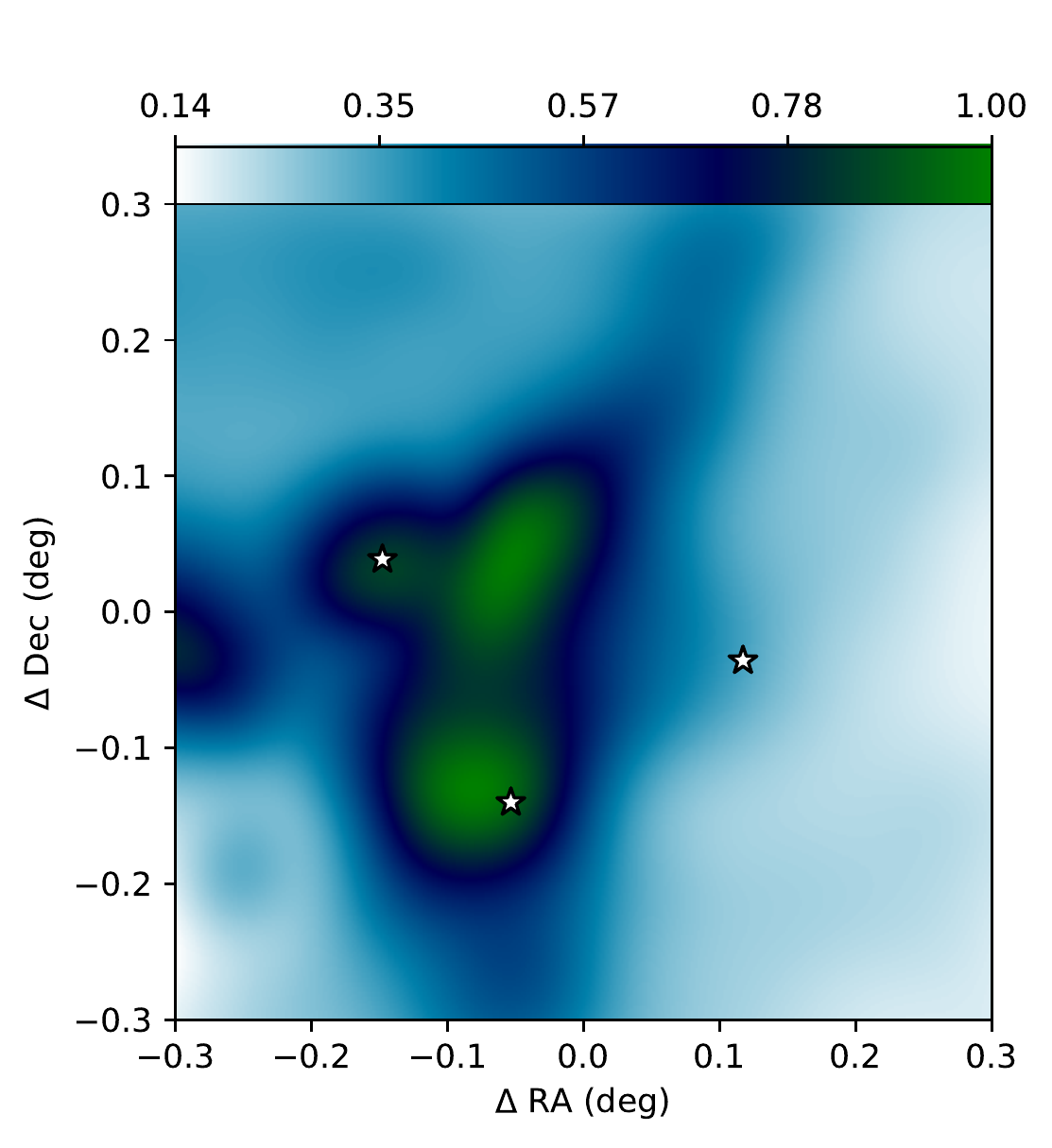}
\caption{I$_{\mathrm{12\,\upmu m}}/G_0$} \label{fig:d}
\end{subfigure}\hspace*{0.5em}
\begin{subfigure}{0.33\textwidth}
\includegraphics[width=\linewidth]{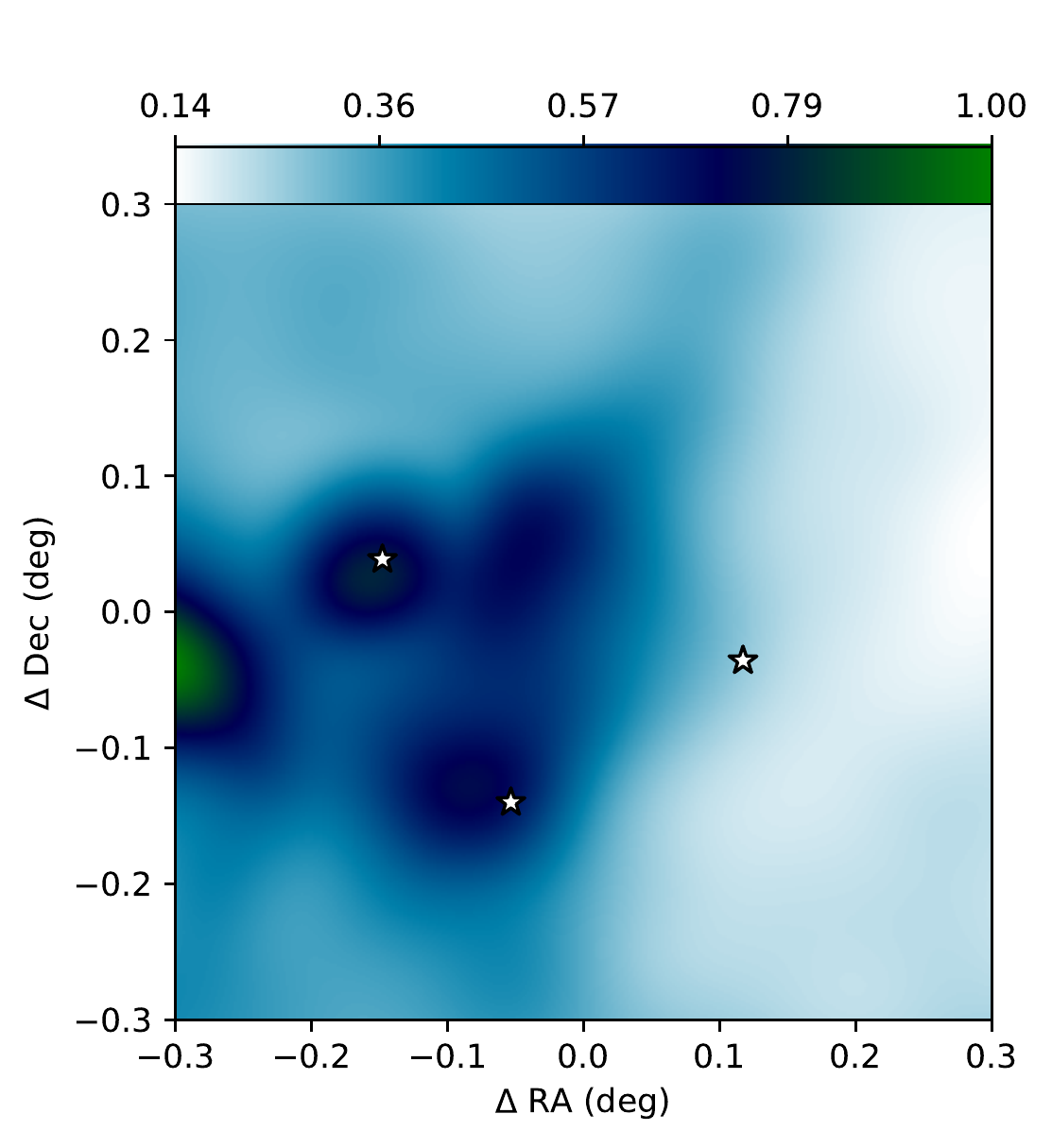}
\caption{I$_{\mathrm{8\,\upmu m}}/G_0$} \label{fig:e}
\end{subfigure}\hspace*{0.5em}
\begin{subfigure}{0.33\textwidth}
\includegraphics[width=\linewidth]{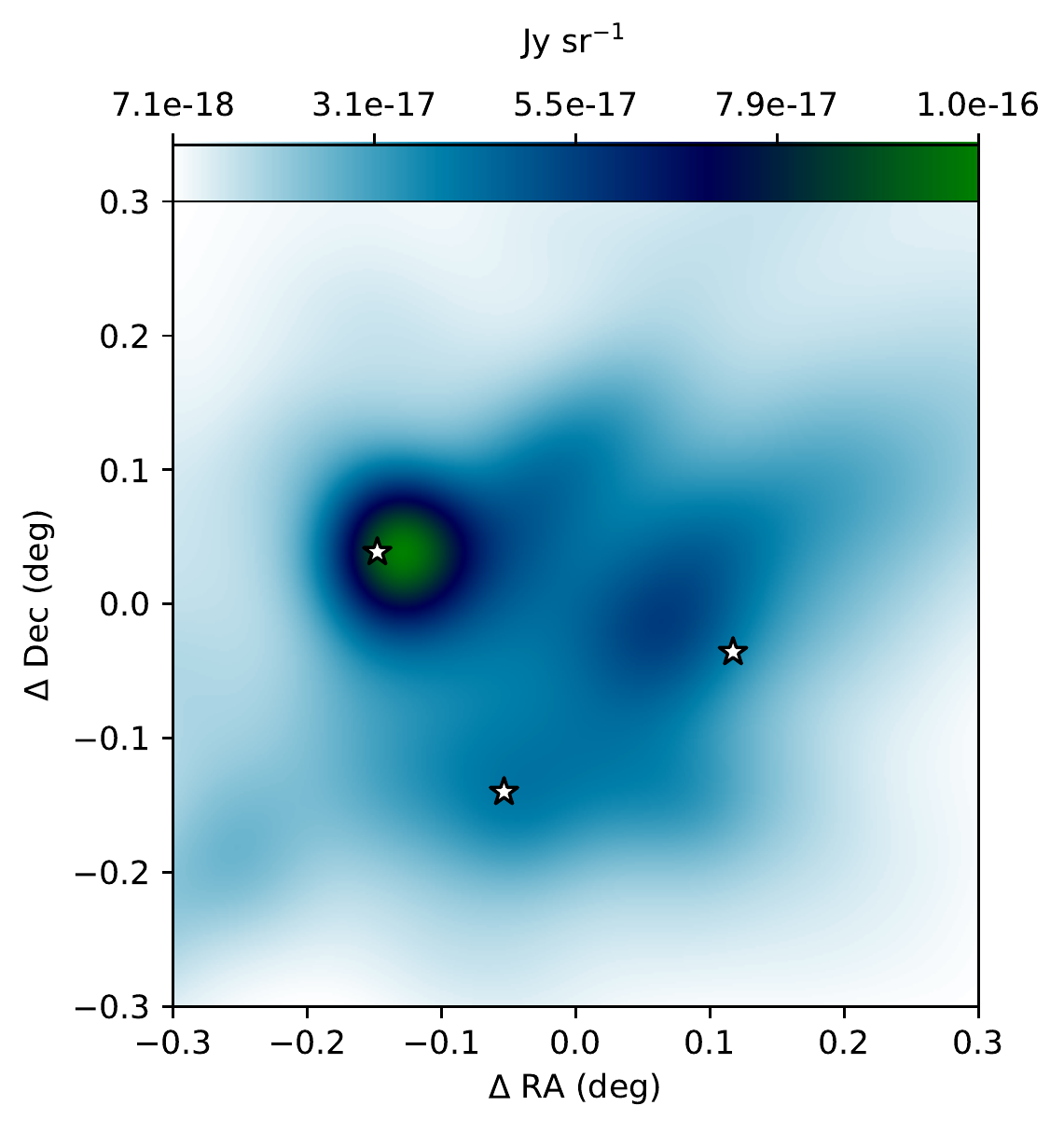}
\caption{$\mathcal{R}$/$\nu$} \label{fig:f}
\end{subfigure}
\caption{{\tt clean} 31\,GHz map (\ref{fig:a}, close-up from Fig.\,\ref{fig:cbi-points}), normalized proxies for PAHs column densities (\ref{fig:b}-\ref{fig:e}) and radiance map (\ref{fig:f}). The proxies for column densities correlate better with the 31 GHz data, and the morphology between them is very similar. In \ref{fig:a}, the dashed region shows the area within which we calculated the plane of the sky correlations; it corresponds to 50$\%$ of the mosaic's primary beam. The position for the three early-type stars are also marked in each map.} \label{fig:correlation-fig}
\end{figure*}

Another way to obtain a proxy for the column of PAHs, proposed by \cite{hensley16}, is to correct I$_{\nu\,\mathrm{PAH}}$ by the  dust radiance ($\mathcal{R}$) as a method to quantify the fraction of dust in PAHs ($f_{\mathrm{PAH}}$). The dust radiance corresponds to the integrated intensity, $\mathcal{R}$ = $\int_{\nu} I_{\nu}$. We calculated this expression by considering a modified blackbody as in \citet[][equation 10]{planck14}: 
\begin{ceqn}
\begin{align}
\mathcal{R} = \tau_{250} \frac{\sigma_{S}}{\pi} \, T^{4}_{\mathrm{dust}} \left(\frac{k T_{\mathrm{dust}}}{h \nu_0} \right)^{\beta_{\mathrm{dust}}}\,\frac{C(4 + \beta_{\mathrm{dust}} )\, \zeta(4+ \beta_{\mathrm{dust}})}{\Gamma(4)\,\zeta(4)},
\end{align}
\end{ceqn}
where $\tau_{\mathrm{250}}$ is
the optical depth at 250\,$\upmu$m, calculated using $\tau_{\mathrm{250}}$\,=\,$I_{\mathrm{250\,\upmu m}} / B_{\mathrm{250\,\upmu m}}(T)$ by assuming an optically-thin environment. Also, $\nu_0$ is the frequency for 250\,$\upmu$m, $\sigma_{S}$ is the Stefan-Boltzmann constant, $k$ is the Boltzmann constant, $h$ is the Planck constant and $\Gamma$ and $\zeta$ are the Gamma and Riemann-zeta functions, respectively. To recover the intensity units of the radiance map, we divide it by $\nu_0$, getting as a result
$\mathcal{R}$/$\nu$. Hence, the PAH fraction can be calculated as:
\begin{ceqn}
\begin{align}
\frac{\nu\,I_{\nu\,\mathrm{PAH}}}{\mathcal{R}} \propto f_{\mathrm{PAH}}.
\end{align}
\end{ceqn}
Thus, the product of the PAH fraction times the optical depth will be proportional to the PAH column density:
\begin{ceqn}
\begin{align}
\label{eq:hensley_proxy}
f_{\mathrm{PAH}} \, \times \, \tau_{250} \propto N_{\mathrm{PAH}} .
\end{align}
\end{ceqn}
We note that \citet{hensley16} stress the need of a good correlation between the mid-IR map and $\mathcal{R}$, which is the case of our 4.5\,arcmin I$_{\nu\,\mathrm{PAH}}$ and $\mathcal{R}$ maps, with correlation coefficients $r$\,>\,0.6. In Figure\,\ref{fig:correlation-fig} we show a close-up of the 31\,GHz {\tt clean} mosaic, along with the column density proxies constructed with $\mathrm{WISE}$\,12\,$\upmu$m and $\mathrm{IRAC}$\,8\,$\upmu$m, and the radiance map.

\begin{figure*} 
\begin{subfigure}{0.45\textwidth}
\includegraphics[width=\linewidth]{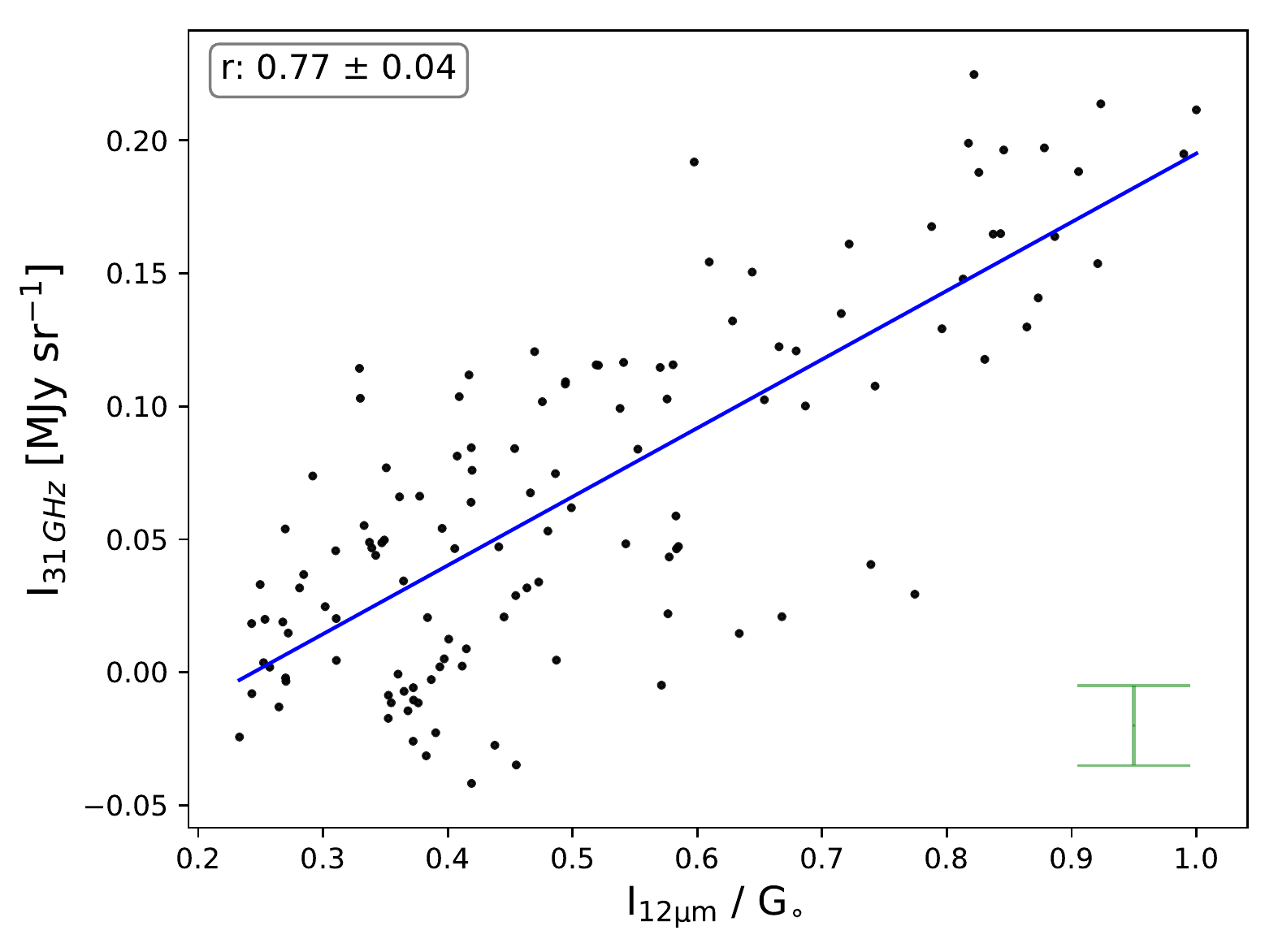}
\caption{ } \label{fig:scatter1}
\end{subfigure}\hspace*{1.8em}
\begin{subfigure}{0.45\textwidth}
\includegraphics[width=\linewidth]{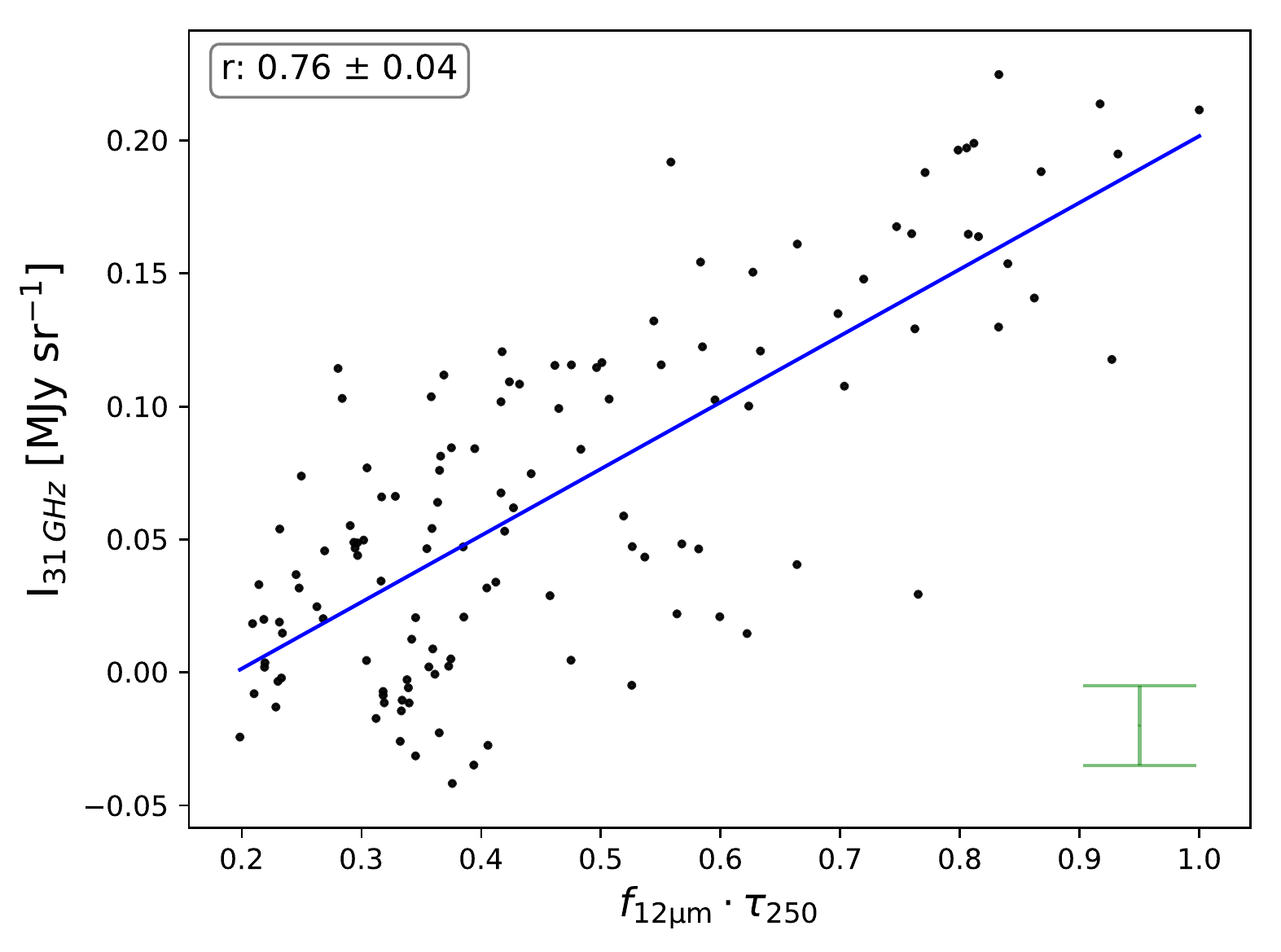}
\caption{ } \label{fig:scatter2}
\end{subfigure}
\caption{Linear correlations between the 31 GHz data and the column density proxies I$_{12\,\mathrm{\upmu m}}$/G$_{\circ}$ (\ref{fig:scatter1}) and f$_{12\,\mathrm{\upmu m}}$ $\cdot$ $\tau_{250}$ (\ref{fig:scatter2}), in the plane of the sky. RMS error bars ($\sim$ 0.03 MJy\,sr$^{-1}$) are shown for the 31\,GHz data. The column density proxies axis (x-axis) are normalized in both cases and their errors are negligible.} 
\label{fig:scatter}
\end{figure*}

\subsection{Pearson correlation analysis}

Figure\,\ref{fig:scatter} shows the scatter plots using the 31\,GHz data and column density proxies I$_{\mathrm{12\,\upmu m}}/G_0$ and $f_{\mathrm{\,12\,\upmu m}} \, \cdot \, \tau_{250}$ in the plane of the sky. These scatter plots suggest a linear dependence between the CBI\,2 data and each template, which justifies the use of the Pearson coefficient as a statistical measurement of the correlation. The Pearson correlation coefficient, $r$, provides a way to quantify and compare the degree of correlation between the 31\,GHz signal and different templates, both in the $uv$-domain,
\begin{equation}
r_{uv} = \frac{\sum_k \omega_k \left(Vx_k - \overline Vx \right) \left(Vy_k - \overline Vy \right)}{\sqrt{\sum_k \omega_k \left(Vx_k - \overline Vx \right)^2 \sum_k \omega_k \left(Vy_k - \overline Vy \right)^2}},
\end{equation}
where $\omega_k$ is the weight for visibility datum $V_k$, and $\overline Vx$\,=\,$\sum_k \omega_k Vx_k/\sum_k \omega_k$ and $\overline Vy$\,=\,$\sum_k \omega_k Vy_k/\sum_k \omega_k$ are the weighted means of visibility data-sets $Vx$ and $Vy$, respectively; and in the sky-plane
\begin{equation}
r_\mathrm{sky} = \frac{\sum_i \left(x_i - \overline x \right) \left(y_i - \overline y \right)}{\sqrt{\sum_i \left(x_i - \overline x \right)^2 \sum_i \left(y_i - \overline y \right)^2}},
\end{equation}
where $\overline x$ and $\overline y$ are the mean values of data-sets $x$ and $y$, respectively.

The correlations in the $uv$-plane ($r_{\mathrm{uv}}$) are calculated for the entire visibility data-set. On the other hand, the sky-plane cross-correlations are taken inside an area equivalent to half the mosaic's primary beam FWHM, shown as a dashed region in Figure\,\ref{fig:a}. This helps us avoid noisy outliers and measure the correlation within the area of interest, which is the PDR. We also masked the PMN galaxy identified in Figure\,\ref{fig:cbi-points}. 

The uncertainties  in $r_{\mathrm{sky}}$ were estimated with a Monte Carlo simulation. We added random Gaussian noise to the 31\,GHz {\tt clean} mosaic, before dividing by the mosaic attenuation map, with a dispersion given by the RMS noise of the 31\,GHz {\tt clean} mosaic. In each run of the simulations, the resulting 31\,GHz intensity map is correlated with the corresponding template, and the final error in $r_{\mathrm{sky}}$ is the standard deviation of all the correlation coefficients. In the $uv$-plane, the $r_{\mathrm{uv}}$ errors are calculated under the same logic, but instead, we add random Gaussian noise to each visibility datum $V_k$ using a dispersion given by its weight $\omega_k = 1/\sigma_k^2$, common to both the real and imaginary parts. We are interested in the relative variations of the Pearson coefficients between the templates, so any bias due to systematic errors will be the same for all the maps. 

The resulting values for the correlations coefficients are listed in Table\,\ref{tab:correl}. We observe the same trends in both $r_{\mathrm{sky}}$ and $r_{\mathrm{uv}}$, although the sky-plane results show higher correlations. This difference is expected, as the sky correlations were extracted within 50\,$\%$ of the primary beam area, while the $uv$-plane correlations used all of the visibilities, which are integrated quantities over the whole primary beams. 

Figure\,\ref{fig:slit_proxies} shows the cuts for the normalized radiance, $\tau_{250}$, and column density proxies, starting from star HD\,147889 and passing through the 31\,GHz emission peak (as shown by the arrow in Figure\,\ref{fig:globalrgb}). Here, we see that the column density proxies of the small grains peak around the same location as the 31\,GHz data. Note the shift of the IRAC\,8$\upmu$m and WISE\,12$\upmu$m profiles between Fig.\,\ref{fig:slit_ir} and their column density proxies in Fig.\,\ref{fig:slit_proxies}. It is also interesting to highlight that the location of the bump in the WISE\,12$\upmu$m profile (at around 10\,arcmin in Fig.\,\ref{fig:slit_ir}) matches the 31\,GHz peak. This leads to a stronger correlation between both templates, which is also reflected in Fig.\,\ref{fig:slit_proxies}, where I$_{\mathrm{12\,\upmu m}}/G_0$ and $f_{\mathrm{\,12\,\upmu m}} \, \cdot \, \tau_{250}$ are narrower and show a more similar profile to the 31\,GHz data. As shown in Table\,\ref{tab:correl}, the best correlation is indeed with PAH column density proxy I$_{\mathrm{12\,\upmu m}}/G_0$. The role of the WISE\,3$\upmu$m map as a tracer for smaller PAHs will be discussed in Sec.\,\ref{sec:pahsize}.

Note that, in Figure\,\ref{fig:slit_proxies}, the radiance profile shows two peaks: one around 5\,arcmin from HD\,147889, which is related to hotter dust grains nearby the star, and a second one along the PDR but shifted $\sim$2\,arcmin from the 31\,GHz profile, which is related to a larger optical depth ($\tau_{250}$). The $\tau_{250}$ peak, which traces the concentration of big grains, is shifted towards the molecular cloud, matching the peak of $\mathrm{SPIRE}$\,250\,$\upmu$m (Fig.\,\ref{fig:slit_ir}) at $\sim$13\,arcmin. Table\,\ref{tab:correl} shows that the $\tau_{250}$ map does not correlate with the 31\,GHz map, while the correlation coefficient for the radiance template ($\mathcal{R}$) is moderate. Historically, sub-mm templates have been used to trace dust emission \citep{dickinson18}, but, in this case, the quantification of the sub-mm emission is better traced by the radiance in comparison with $\tau_{250}$ or the intensity at any given sub-mm frequency. This is consistent with the result by \citet{hensley16} where the radiance template is the best tracer for the AME.

\begin{table}
        \centering
        \caption{Pearson correlations between the 31\,GHz data and different templates, for the plane of the sky ($r_{\mathrm{sky}}$) and the $uv$-plane ($r_{\mathrm{uv}}$).}
        \label{tab:correl}
        \begin{tabular}{ccc} 
               \hline
Template & {$r_{\mathrm{\,uv}}$} & {$r_{\mathrm{\,sky}}$}\\
                \hline
WISE\,3  & 0.25 $\pm$ 0.01 & 0.61 $\pm$ 0.04  \\
IRAC\,8  & 0.26 $\pm$ 0.01 & 0.68 $\pm$ 0.04  \\
WISE\,12 & 0.29 $\pm$ 0.01 &  0.72 $\pm$ 0.04  \\
WISE\,22 & 0.15 $\pm$ 0.01 &  0.52 $\pm$ 0.04  \\
MIPS\,24 & 0.15 $\pm$ 0.01 &  0.51 $\pm$ 0.04  \\
SPIRE\,250 & 0.08 $\pm$ 0.01  & 0.40 $\pm$ 0.04  \\
$\tau_{250}$ & -0.04 $\pm$ 0.01 & 0.11 $\pm$ 0.04 \\

$\mathcal{R}$/$\nu$ & 0.13 $\pm$ 0.01 & 0.51 $\pm$ 0.04 \\

$f_{\mathrm{\,3\,\upmu m}}$ & 0.17 $\pm$ 0.01 & 0.23 $\pm$ 0.04 \\
$f_{\mathrm{\,8\,\upmu m}}$ & 0.23 $\pm$ 0.01 & 0.32 $\pm$ 0.02 \\
$f_{\mathrm{\,12\,\upmu m}}$ & 0.25 $\pm$ 0.01 &  0.37 $\pm$ 0.02 \\      

$f_{\mathrm{\,3\,\upmu m}} \, \cdot \, \tau_{250}$  & 0.27 $\pm$ 0.01 &  0.70 $\pm$ 0.04 \\
$f_{\mathrm{\,8\,\upmu m}} \, \cdot \, \tau_{250}$  & 0.28 $\pm$ 0.01 &  0.59 $\pm$ 0.04 \\
$f_{\mathrm{\,12\,\upmu m}} \, \cdot \, \tau_{250}$ & 0.34 $\pm$ 0.01 & 0.76 $\pm$ 0.04 \\

I$_{\mathrm{3\,\upmu m}}/G_0$ & 0.29 $\pm$ 0.01 &  0.71 $\pm$ 0.04 \\
I$_{\mathrm{8\,\upmu m}}/G_0$ & 0.26 $\pm$ 0.01 &  0.60 $\pm$ 0.04 \\
I$_{\mathrm{12\,\upmu m}}/G_0$ & 0.34 $\pm$ 0.01 & 0.77 $\pm$ 0.04 \\

              \hline
\end{tabular} 
\end{table}

\begin{table}
        \centering
        \caption{Pearson coefficients for the original visibility data-set and the visibility data-set using uv-tapering for an equivalent angular resolution of 13.5\,arcmin ($\sim$\,3 times the original CBI\,2 angular resolution). We show the ratio between the Pearson coefficients of the two data-sets. In particular, we see that the dust radiance coefficient tends to increase the most at a lower equivalent angular resolution.}
        \label{tab:suavizado}
        \begin{tabular}{ccc|c} 
               \hline
Template & {$r_{\mathrm{\,uv}}$} & {$r_{\mathrm{\,uv-13.5\,'}}$ } & {Ratio} \\
                \hline
WISE\,3  & 0.25 $\pm$ 0.01 & 0.32 $\pm$ 0.02        &  1.3 $\pm$ 0.1\\
IRAC\,8  & 0.26 $\pm$ 0.01 & 0.35 $\pm$ 0.02                        &  1.3 $\pm$ 0.1\\
WISE\,12 & 0.29 $\pm$ 0.01 & 0.37 $\pm$ 0.02                        & 1.3 $\pm$ 0.1  \\
WISE\,22 &  0.15 $\pm$ 0.01 & 0.20 $\pm$ 0.01                       &  1.4 $\pm$ 0.1 \\
MIPS\,24 & 0.15 $\pm$ 0.01 & 0.20 $\pm$ 0.02                        & 1.3 $\pm$ 0.2 \\
SPIRE\,250 & 0.08 $\pm$ 0.01 & 0.09 $\pm$ 0.01                       &  1.1 $\pm$ 0.2\\
$\tau_{250}$ & -0.04 $\pm$ 0.01 & -0.06 $\pm$ 0.01                  &  1.5 $\pm$ 0.4\\
$\mathcal{R}$/$\nu$ & 0.13 $\pm$ 0.01 & 0.20 $\pm$ 0.02             & 1.5 $\pm$ 0.2 \\

$f_{\mathrm{\,3\,\upmu m}}$ & 0.17 $\pm$ 0.01  & 0.21  $\pm$ 0.02 & 1.2 $\pm$ 0.1  \\ 
$f_{\mathrm{\,8\,\upmu m}}$ & 0.23 $\pm$ 0.01 & 0.29 $\pm$ 0.02     & 1.3 $\pm$ 0.1 \\
$f_{\mathrm{\,12\,\upmu m}}$ & 0.25 $\pm$ 0.01  &  0.31 $\pm$ 0.02 & 1.2 $\pm$ 0.1  \\ 

$f_{\mathrm{\,3\,\upmu m}} \, \cdot \, \tau_{250}$  & 0.27 $\pm$ 0.01 & 0.34 $\pm$ 0.01 & 1.3 $\pm$ 0.1\\
$f_{\mathrm{\,8\,\upmu m}} \, \cdot \, \tau_{250}$  & 0.28 $\pm$ 0.01 & 0.37 $\pm$ 0.03 & 1.3 $\pm$ 0.1\\
$f_{\mathrm{\,12\,\upmu m}} \, \cdot \, \tau_{250}$  & 0.34 $\pm$ 0.01 & 0.42 $\pm$ 0.03 & 1.2 $\pm$ 0.1\\

I$_{\mathrm{3\,\upmu m}}/G_0$ & 0.29 $\pm$ 0.01 & 0.37 $\pm$ 0.03   &  1.3 $\pm$ 0.1 \\
I$_{\mathrm{8\,\upmu m}}/G_0$ & 0.26 $\pm$ 0.01 & 0.34 $\pm$ 0.03   &  1.3 $\pm$ 0.1 \\
I$_{\mathrm{12\,\upmu m}}/G_0$ &  0.34 $\pm$ 0.01 & 0.42 $\pm$ 0.03 & 1.2 $\pm$ 0.1 \\

              \hline
\end{tabular} 
\end{table}

\subsection{Correlation analysis as a function of angular resolution}
The cross-correlations also depend on angular resolution. We repeated the correlations in the $uv$-plane for an equivalent angular resolution of 13.5\,arcmin, closer to the approximate maximum recoverable scale ($\sim$\,three times the CBI\,2 synthesized beam). To do this, we applied a $uv$-taper by multiplying the visibility weights with a Gaussian, $W$\,=\,$\exp{\left(-(u^2 + v^2)/t^2  \right)}$. The results are presented in Table\,\ref{tab:suavizado}, where we also show the ratio between the Pearson coefficients of the original and the tapered data-set. These correlations show the highest increase for $\mathcal{R}/\nu$ and $\tau_{\mathrm{250}}$, but note that the correlation for $\tau_{\mathrm{250}}$ is very close to zero. We also calculated the correlations in the sky plane by smoothing the templates to angular resolutions of 9 and 13.5\,arcmin and observed the same tendency: the Pearson coefficient of the dust radiance maps increases the most at lower angular resolutions, while the column density proxies tend to remain equal.

The fact that the radiance template shows the highest relative increase in its Pearson coefficient could explain the results obtained by \citet{hensley16}, where their dust radiance template correlated the best with their AME template at an angular resolution of 1$^{\circ}$. Based on their correlation results, they concluded that PAHs might not be responsible for the AME. This is not the case for our 4.5\,arcmin maps, so we believe that PAHs cannot be ruled out as possible SD carriers. In this PDR, the 8\,$\upmu$m and 12\,$\upmu$m bands are dominated by PAHs \citep[see e.g.,][]{habart03, drli07, compiegne+11}, so the fact that we found a good correlation between these bands and the 31\,GHz emission indicates that the PAHs might be important AME emitters. 

\begin{table*}
	\centering
		\begin{threeparttable}
	\caption{AME correlation slopes for $\rho$\,Oph\,W between typical AME frequencies and dust templates. The $\tau_{353}$ map can be considered a better dust template as it traces the dust column density. We find that most of the $\rho$\,Oph 31\,GHz flux must be coming from $\rho$\,Oph\,W.}
	\label{tab:emissivities}
	\begin{tabular}{cccccccc} 
		\hline
& {Frequency [GHz]} & Angular Resolution &  {Correlation slope} & {Units} & {Template} & {Reference} & \\
        \hline
& 31 & 4.5\,arcmin & 15.5 $\pm$ 0.2  & 10$^{6} \upmu$K & $\tau_{353}$ & This work. & \\
& 22.8 & 1$^{\circ}$ & 23.9 $\pm$ 2.3  & 10$^{6} \upmu$K & $\tau_{353}$ & \citet{dickinson18} & \\

& 31 & 4.5\,arcmin & 2.18 $\pm$ 0.04  & $\upmu$K\,/\,(MJy sr$^{-1}$) & 100\,$\upmu$m (IRIS) & This work. & \\
& 22.8 &1$^{\circ}$ & 8.3 $\pm$ 1.1  & $\upmu$K\,/\,(MJy sr$^{-1}$) & 100\,$\upmu$m (IRIS) & \citet{dickinson18} & \\
		\hline
	\end{tabular}
    	\end{threeparttable}
\end{table*}

\subsection{AME correlation slopes}

Typically, AME emissivities have been quantified with correlation slopes in terms of IR emission (e.g. I$_{\mathrm{30\,GHz}}$ / I$_{\mathrm{100\,\upmu m}}$) or using an optical depth map (I$_{\mathrm{30\,GHz}}$ / $\tau_{\mathrm{353\,GHz}}$) \citep[see][]{dickinson18}. We calculated the correlation slope between our 31\,GHz data, and $\tau_{353}$ and IRIS\,100\,$\upmu$m. For this, we measured the mean emission within the PDR filament using intensities larger than 15$\sigma$ (as the contour defined in Fig.\ref{fig:cbi-points}) in the primary beam corrected {\tt clean} map. The original angular resolutions for the $\tau_{353}$ and IRIS\,100\,$\upmu$m maps (5 and 4.3\,arcmin, respectively) are very similar to CBI\,2's angular resolution of 4.5\,arcmin. In Table\,\ref{tab:emissivities} we list the values for the AME correlation slopes. We also show the correlation slopes measured with the same templates by \citet{dickinson18}, at an angular resolution of 1$^{\circ}$. We find that a large fraction of the AME correlation slope measured in $\rho$\,Oph on scales of 1$^{\circ}$ must be coming from the $\rho$\,Oph\,W PDR. This is especially evident when using the $\tau_{353}$ map, where the AME correlation slope of the 4.5\,arcmin $\rho$\,Oph\,W PDR observations is $\sim$65$\%$ of the AME correlation slope of the 1\,$^{\circ}$ $\rho$\,Oph observations. $\rho$\,Oph\,W shows the highest AME emissivity in terms of the $\tau_{353}$ map, which is considered as a very reliable dust template as it traces the dust column density \citep{dickinson18}. These emissivities are about a factor 2-3 higher than other values measured at high Galactic latitudes and also in Perseus, the brightest AME source in the sky according to \cite{planck}. A possible explanation for the larger slope in $\rho$\,Oph\,W could be found in the finer linear resolution in this work, of $\sim$0.2\,pc, compared to $\sim$2.3\,pc in \cite{planck}. The fact that $\rho$\,Oph is closer to us means that we can resolve the AME from the PDR. When integrated in the telescope beam (e.g. Planck analysis over 1$^{\circ}$ scales), there is proportionally a larger amount of AME coming from the high density and excited PDR compared to the emission from the regions that have lower emissivities, i.e. AME from the denser cores, or from the more diffuse gas.

\begin{figure}
\begin{subfigure}{\columnwidth}
\includegraphics[width=\linewidth]{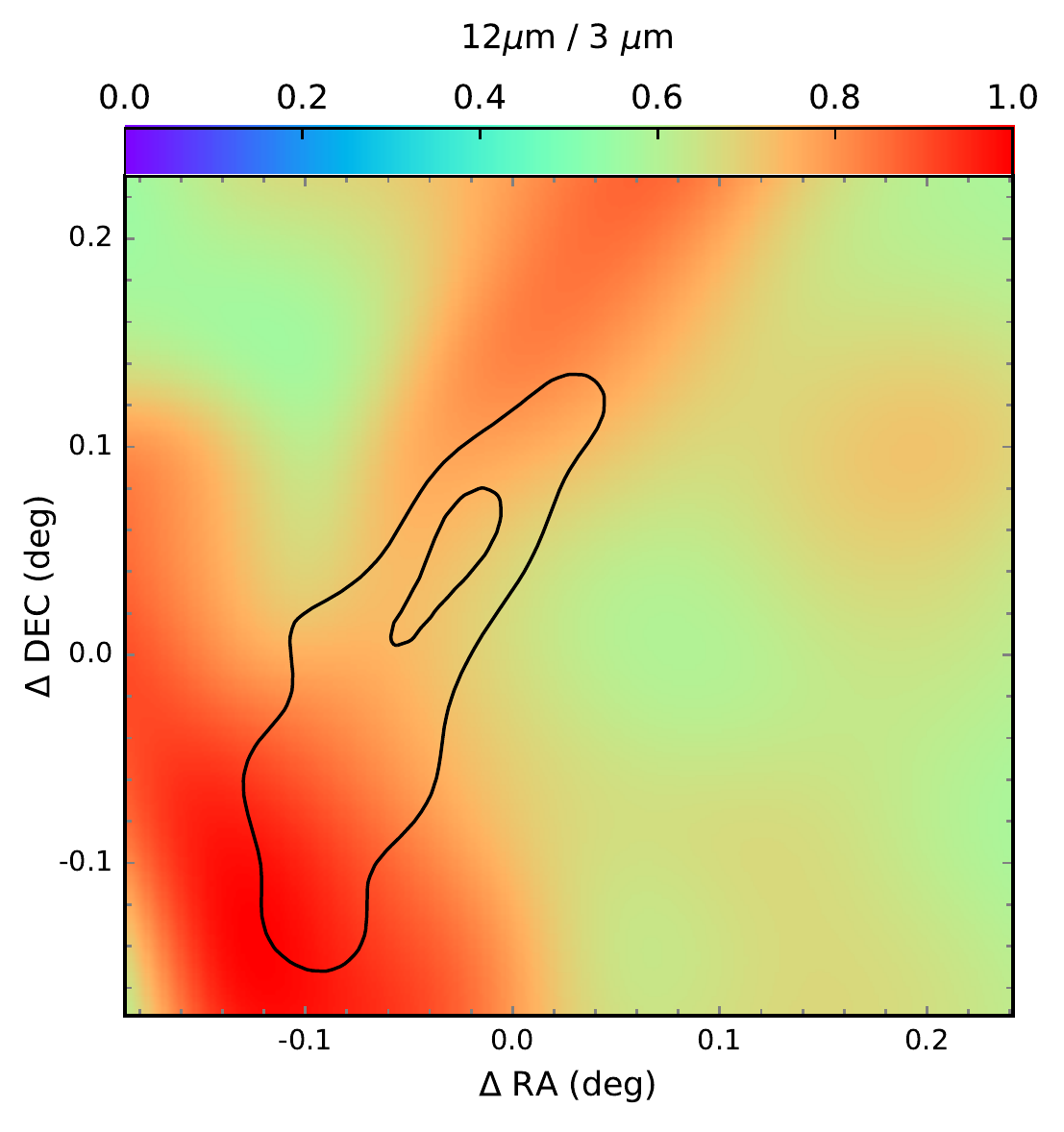}
\caption{} \label{fig:12/3}
\end{subfigure}
\medskip
\begin{subfigure}{\columnwidth}
\includegraphics[width=\linewidth]{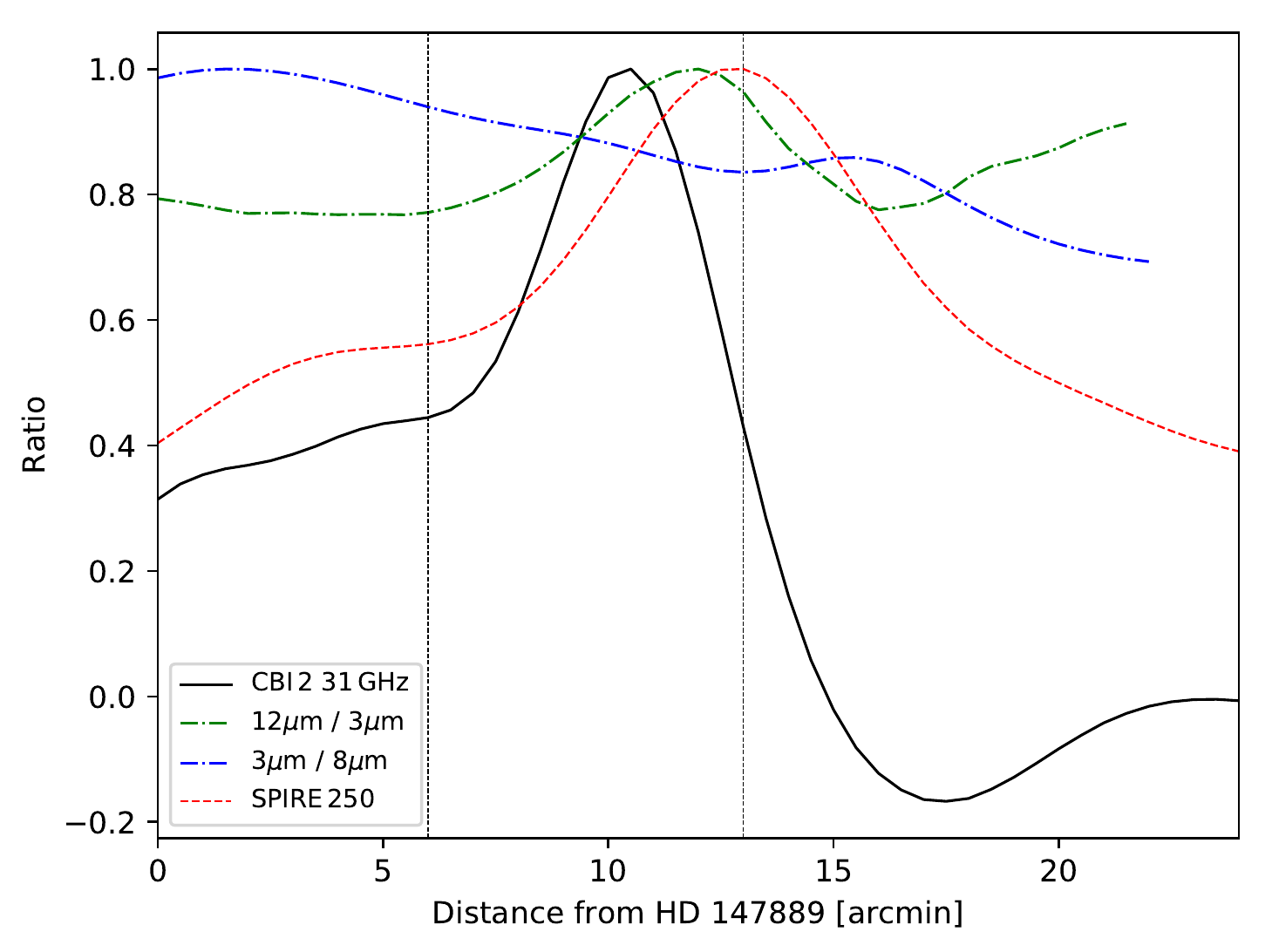}
\caption{} \label{fig:cut_ratio}
\end{subfigure}
\caption{\ref{fig:12/3}: Normalized 12\,$\upmu$m\,/\,3\,$\upmu$m ratio in the $\rho$\,Oph\,W PDR. \ref{fig:cut_ratio}: normalized emission profiles vs. distance to ionizing star HD\,147889 for the 12\,$\upmu$m\,/\,3\,$\upmu$m and 3\,$\upmu$m\,/\,8\,$\upmu$m ratios in comparison to the 31\,GHz and 250\,$\upmu$m emission. The profile cut corresponds to the dashed arrow in Fig.\,\ref{fig:globalrgb} and the black vertical lines define the PDR.} \label{f:pah_size_ratio}
\end{figure}

\begin{figure*}
\begin{subfigure}{0.33\textwidth}
\includegraphics[width=\linewidth]{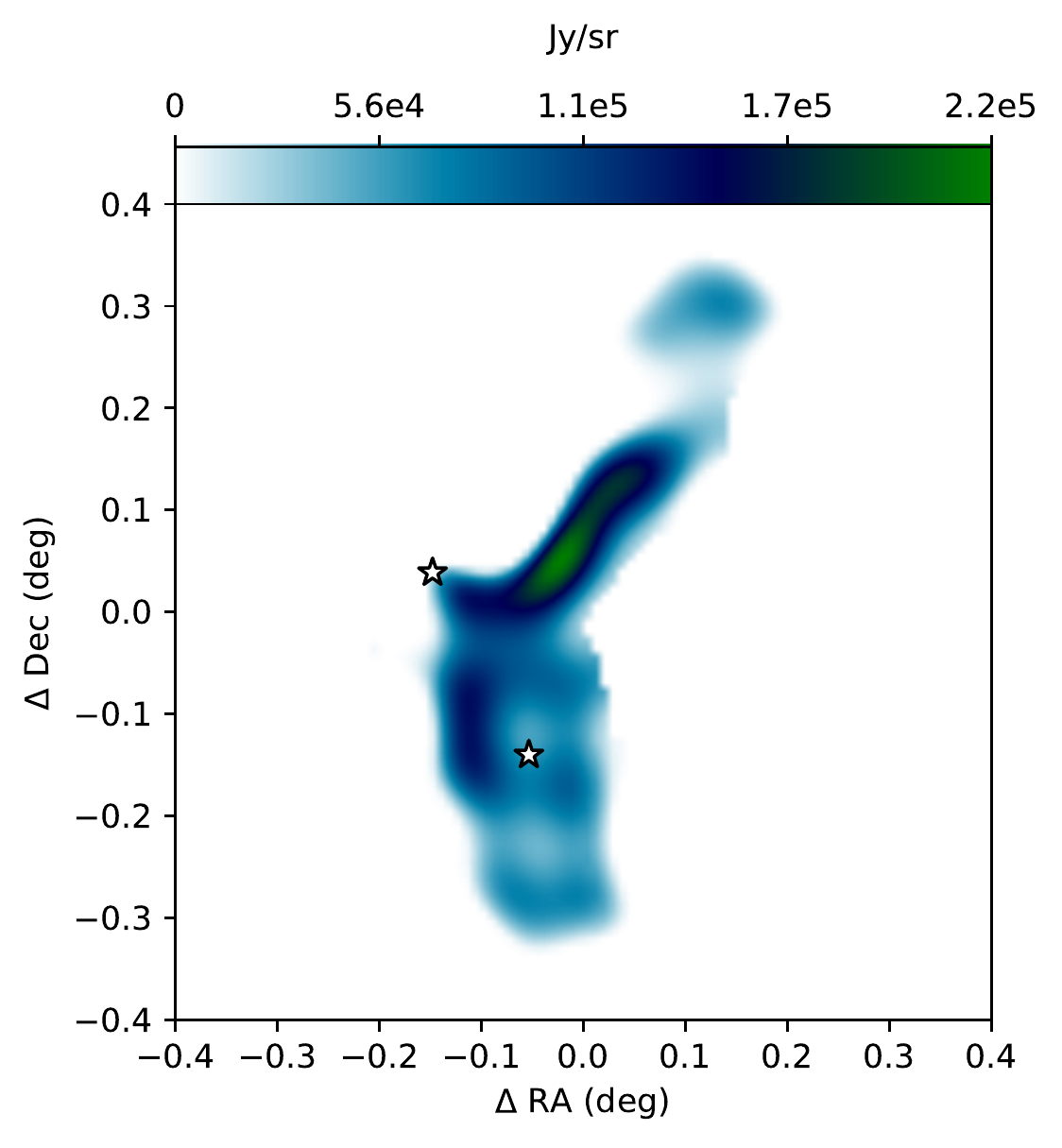}
\caption{{\tt gpu-uvmem} model map} \label{fig:a-rmap}
\end{subfigure}\hspace*{0.5em}
\begin{subfigure}{0.33\textwidth}
\includegraphics[width=\linewidth]{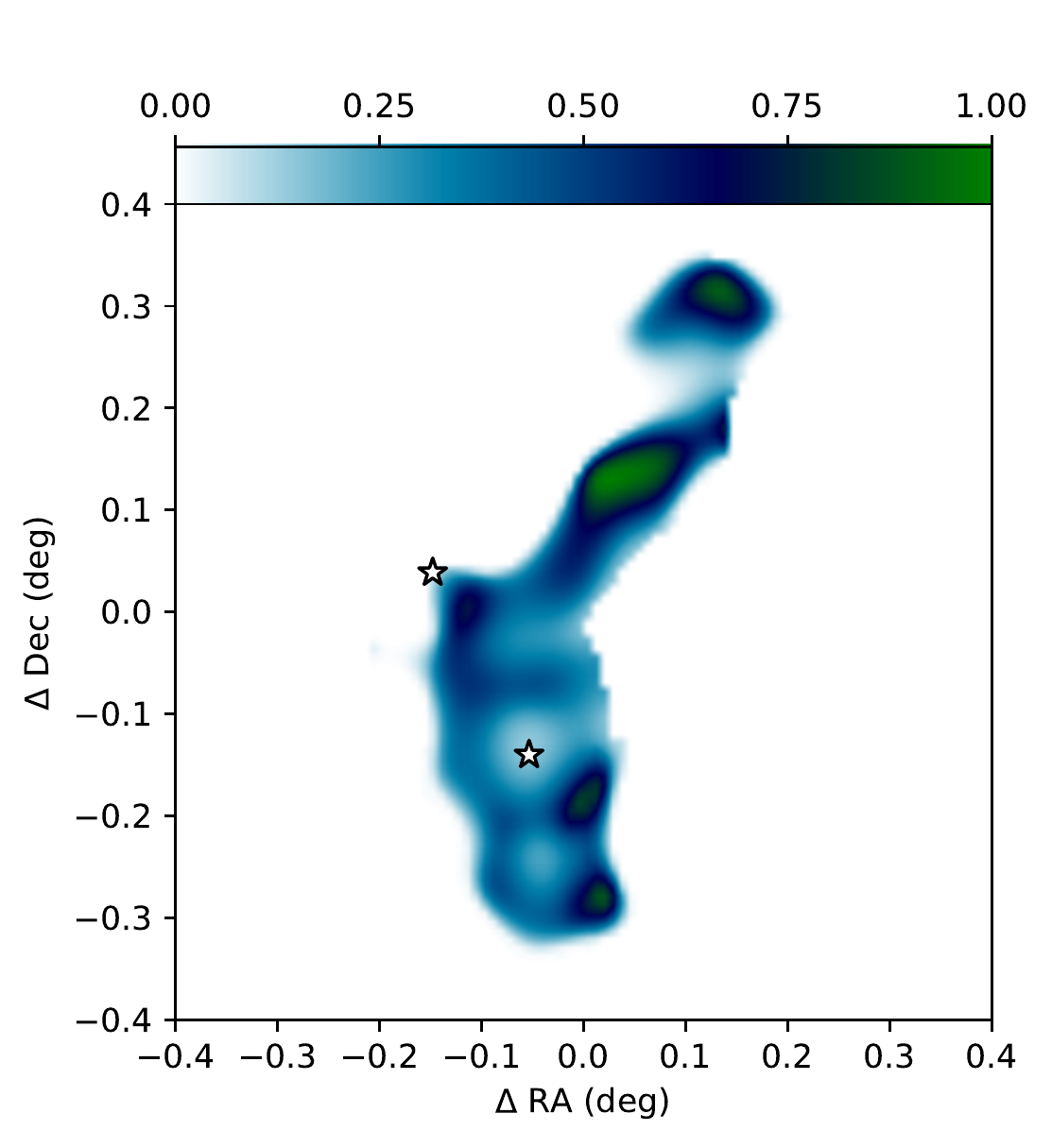}
\caption{{\tt gpu-uvmem} $\hat{j_{\nu}}_{\,\mathrm{proxy}}$ map} \label{fig:b-rmap}
\end{subfigure}\hspace*{0.5em}
\begin{subfigure}{0.33\textwidth}
\includegraphics[width=\linewidth]{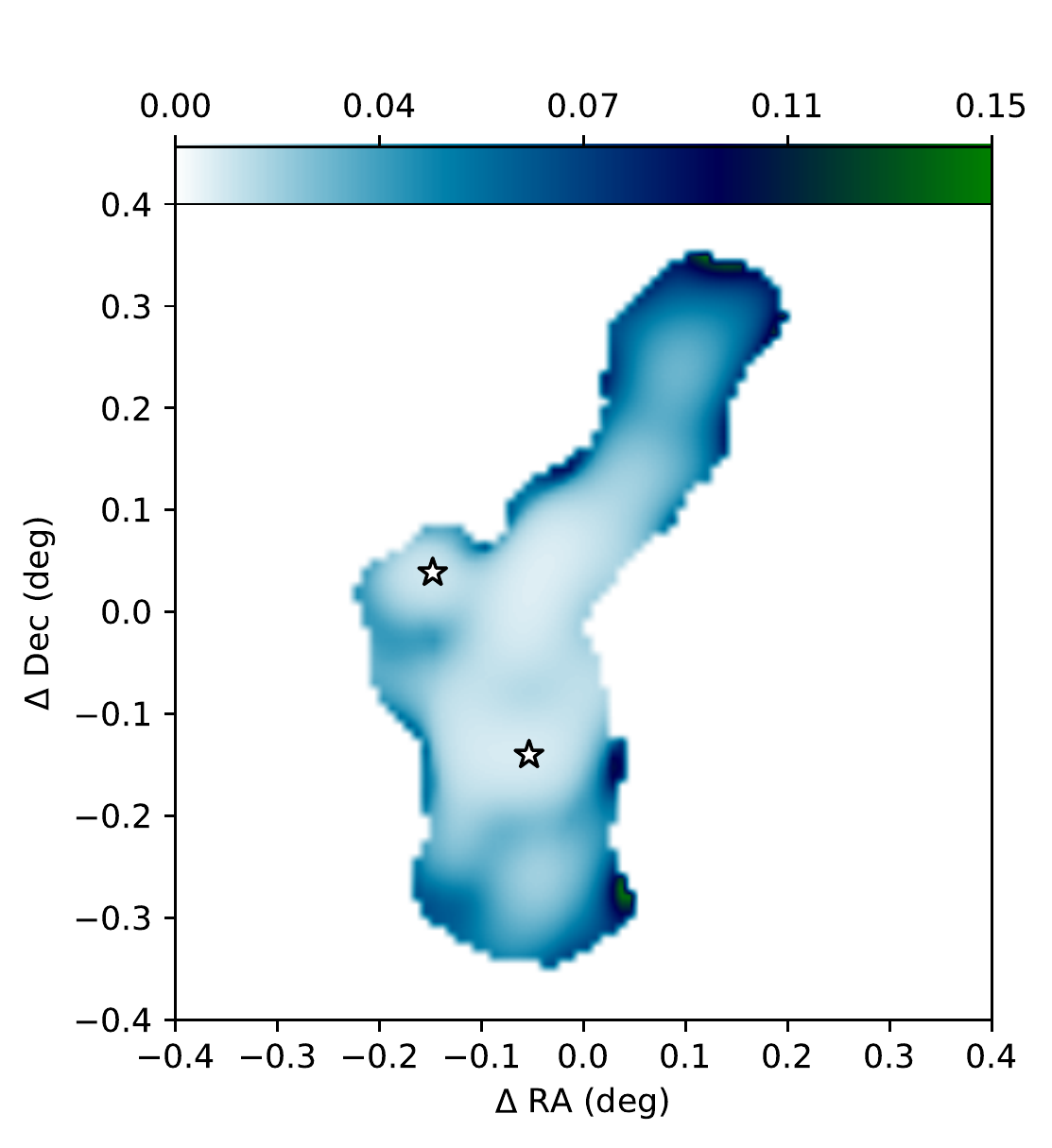}
\caption{$\hat{j_{\nu}}_{\,\mathrm{proxy}}$ noise map} \label{fig:c-rmap}
\end{subfigure}
\caption{\ref{fig:a-rmap}: CBI\,2 31\,GHz {\tt gpu-uvmem} model image. \ref{fig:b-rmap}: normalized $\hat{j_{\nu}}_{\,\mathrm{proxy}}$ map (with a peak of 2.19 $\pm$ 0.04, in arbitrary units) and \ref{fig:c-rmap}: $\hat{j_{\nu}}_{\,\mathrm{proxy}}$\,-\,normalized noise map (the scale corresponds to that of Fig.\,\ref{fig:b-rmap}). We used a mask around HD\,147889 in order to avoid the bulk of the free-free emission. The observed blob towards the northern part of the filament is detected at 3$\sigma$ over the noise in the CBI\,2 {\tt clean} map and could be the continuation of the PDR emission (no catalogued source was identified in that position).} 
\label{fig:rmap}
\end{figure*}

\subsection{Spectral index}
\label{sec:specind}
The wide frequency coverage of the CBI\,2 correlator, between 26\,GHz and 36\,GHz, may place constraints on the spectral properties of the 31\,GHz signal. We split the visibilities ($V$) in two sub-sets: one for the low frequency channels (6-10) centered at 28.5\,GHz, and one for the high frequency channels (1-5) centered at 33.5\,GHz. We then calculated the correlation slopes (a$^{\mathrm{High}}$, a$^{\mathrm{Low}}$) between the two sub-sets and the mock visibilities for I$_{\mathrm{12\,\upmu m}}/G_0$ (the best correlated proxy), following:
\begin{ceqn}
\begin{align}
     V_{31\,\mathrm{GHz}}^{\mathrm{High/Low}} = \mathrm{a}^{\mathrm{High/Low}} \,\,V_{12\,\upmu \mathrm{m}/G_{0}}.
	\label{eq:specind}
\end{align}
\end{ceqn}
Note that $\log{(\mathrm{a}^{\mathrm{High}}/\mathrm{a}^{\mathrm{Low}})}$ is proportional to the spectral index $\alpha$. We obtained a spectral index of $\alpha$\,= 0.05\,$\pm$\,0.08, which indicates a flat spectrum. However, we must also take into account the difficulty of obtaining a reliable spectral index due to any possible calibration and systematic errors. The spectral index measured between the low and high frequency channel is nonetheless consistent with the shape of the $\rho$\,Oph spectrum in \citet{planck}.

\section{Qualitative analysis of the PAH sizes}
\label{sec:pahsize}
The results in Table\,\ref{tab:correl} show that column density proxies of PAH tracers at 3\,$\upmu$m, 8\,$\upmu$m and 12\,$\upmu$m correlate the best with the 31\,GHz emission. Note that, in particular, the 3\,$\upmu$m band is dominated by the smallest PAHs (sizes $\sim$0.4\,nm), which are the most prone to destruction by UV radiation, while the 8\,$\upmu$m and 12\,$\upmu$m bands, in comparison, are dominated by bigger PAHs. Given this, a good tracer of the PAH sizes is the 12\,$\upmu$m\,/\,3\,$\upmu$m ratio map \citep{allamandola+85, ricca+12, croiset+16}. To construct it, we removed the point sources in the WISE\,3\,$\upmu$m map, and smoothed both the WISE 12\,$\upmu$m and 3\,$\upmu$m maps to CBI\,2's angular resolution. In Fig.\,\ref{fig:12/3} we show the 4.5\,arcmin map for the 12\,$\upmu$m\,/\,3\,$\upmu$m ratio, along with the 31\,GHz contours (same contours as in Fig.\,\ref{fig:cbi-points}). At this angular resolution, it is interesting to note the increment of the ratio in the transition between the PDR and the molecular cloud to the East in Fig.\,\ref{fig:12/3}. This is also seen in Fig.\,\ref{fig:cut_ratio}, where we show the normalized emission profiles for the cut in Fig.\,\ref{fig:globalrgb} vs. the distance to the ionizing star HD\,147889. Here, the minimum of the 12\,$\upmu$m\,/\,3\,$\upmu$m ratio is aligned with the PDR surface (defined by the first black vertical line, $\sim$6\,arcmin in Fig.\,\ref{fig:cut_ratio}), while the peak of the ratio occurs at the transition between the PDR and the molecular cloud (defined by the second black vertical line, $\sim$14\,arcmin in Fig.\,\ref{fig:cut_ratio}). This peak might be an indication that the PAHs size increases in the PDR towards the molecular cloud.

The 3\,$\upmu$m\,/\,8\,$\upmu$m ratio map can also give us an indication of the contribution of the smallest PAHs. We calculated the plane of the sky correlation between the 3\,$\upmu$m\,/\,8\,$\upmu$m ratio and the 31\,GHz emission, and obtained a Pearson coefficient of $r_{\mathrm{sky}}$\,=\,$-$0.21\,$\pm$\,0.03. The low value of the anti-correlation indicates that the smallest grains do not necessarily correlate better with the 31\,GHz map, as it can also be seen in Table\,\ref{tab:correl}, where the correlation in WISE\,3 is not better than with IRAC\,8. Fig.\,\ref{fig:cut_ratio} also shows the normalized profile for the 3\,$\upmu$m\,/\,8\,$\upmu$m ratio. The behaviour of this template is similar to the 12\,$\upmu$m\,/\,3\,$\upmu$m map, but shifted towards the interior of the molecular cloud. Overall, these ratios reveal a factor of $\sim$1.5 variation in the size of the emitting PAHs throughout the $\rho$\,Oph\,W PDR.

\section{Emissivity Variations}
\label{sec:emiss}

In our correlation analysis (Sec.\,\ref{sec:correl}) we found that the 31\,GHz map correlates best with I$_{\mathrm{12\,\upmu m}}/G_0$, a template for the PAH column density. Also, we note that there is significant scatter about the linear cross-correlations between the 31\,GHz emission and the PAH column density proxies, as shown in Fig.\,\ref{fig:scatter}. Motivated by this, we study the possibility that this scatter is produced by SD emissivity variations along the PDR.

In order to assess the extent to which the physical mechanisms leading to the 31\,GHz emission depend on the local environment, we tested for the hypothesis that the emergent 31\,GHz intensity is \textit{only} proportional to the column of the emitters, i.e. $N_{\mathrm{PAH}}$. We construct a proxy that will let us measure any emissivity variation in the PDR as in \citet{cas08}, and define the 31\,GHz emissivity per SD emitter (i.e. PAHs in our hypothesis), $\hat{j_{\nu}}$, which should be proportional to the SD emissivity per H-nucleus for a fixed grain population:
\begin{ceqn}
\begin{align}
      \hat{j_{\nu}} \equiv  \frac{I_\nu(31\mathrm{GHz})}{N_{\mathrm{PAH}}} \propto j_{\nu} \equiv  \frac{I_{\nu}}{N_{H}}.
	\label{eq:emissivity}
\end{align}
\end{ceqn}
Using Eq.\,\ref{colcass} we can propose a proxy for $\hat{j_{\nu}}$:
\begin{ceqn}
\begin{align}
      {\hat{j_{\nu}}}_{\,\mathrm{proxy}} \equiv  \frac{I_\nu(31\mathrm{GHz})}{I_{\mathrm{12\,\upmu m}}/G_0}.
	\label{eq:emissivity-proxy}
\end{align}
\end{ceqn}

The {\tt gpu-uvmem} model images for $I^\mathrm{uvmem}_\nu(31\mathrm{GHz})$ and $I^\mathrm{uvmem}_{\mathrm{12\,\upmu m}}/G_0$ provide a finer angular resolution than the {\tt clean} mosaic, and may pick up larger variations in emissivity. Figure\,\ref{fig:rmap} shows $I^\mathrm{uvmem}_\nu(31\mathrm{GHz})$ (Fig.\,\ref{fig:a-rmap}), as well as the normalized $\hat{j_{\nu}}_{\,\mathrm{proxy}}$ map defined in Eq.\,\ref{eq:emissivity-proxy} (Fig.\,\ref{fig:b-rmap}), along with its noise map (Fig.\,\ref{fig:c-rmap}). As explained in Section\,\ref{sec:morph}, the bulk of the free-free emission is located around HD\,147889. Given this, we masked a region with a radius of 6\,arcmin around HD\,147889 in the {\tt gpu-uvmem} maps, in order to quantify only the SD component using $\hat{j_{\nu}}_{\,\mathrm{proxy}}$. Note that the $\hat{j_{\nu}}_{\,\mathrm{proxy}}$ map (Fig.\,\ref{fig:b-rmap}) shows variations throughout the region, with a peak that matches the peak at 31\,GHz in the $\rho$\,Oph\,W PDR.

\begin{table}
	\centering
	\caption{Emissivities variations in $\rho$\,Oph measured with the $\hat{j_{\nu}}_{\,\mathrm{proxy}}$ map.}
	\label{tab:emiss_var}
		\begin{threeparttable}
	\begin{tabular}{c c c c c} 
		\hline
	& Ratio & & R &\\
		\hline
	& $\mathrm{S/N}_{\mathrm{\,Peak \,at\,\,3\sigma}}$\tnote{a} &  & 25.6 &\\
	& $\hat{j_{\nu}}_{\mathrm{proxy-\,Peak}}$ / $\hat{j_{\nu}}_{\mathrm{proxy-\,S1}}$\tnote{b} &  & 22.8 $\pm$ 5.2 &\\
	& $\hat{j_{\nu}}_{\mathrm{proxy-\,Peak}}$ / $\hat{j_{\nu}}_{\mathrm{proxy-\,SR3}}$\tnote{c} &  & 6.2 $\pm$ 0.3 &\\
		\hline
	\end{tabular}
		\end{threeparttable}
	\begin{minipage}{\linewidth}\small
	\vspace{5pt}
	 $^\mathrm{a}$\,Ratio between the emissivity peak and its 3$\sigma$ noise.\\
	$^\mathrm{b}$\,Ratio between the peak and at the location of S1.\\
	$^\mathrm{c}$\,Ratio between the peak and at the location of SR3.
	\end{minipage}
\end{table}

In Table\,\ref{tab:emiss_var} we list $\hat{j_{\nu}}_{\,\mathrm{proxy}}$ variations for different locations in the region. We found that the $\hat{j_{\nu}}_{\,\mathrm{proxy}}$ map has a peak that is 25 times higher than the 3$\sigma$ noise. We repeated this calculation on a {\tt clean} $\hat{j_{\nu}}_{\,\mathrm{proxy}}$ map and found an emissivity peak 10 times stronger than its 3$\sigma$ noise. This difference is expected, as the {\tt gpu-uvmem} map resolves much better the CBI\,2 morphology. 

Interestingly, the emissivity variation between the $\hat{j_{\nu}}_{\,\mathrm{proxy}}$ peak and the location of S1 is 22.8, very close to the strongest variation in $\rho$\,Oph\,W at 3$\sigma$. The fact that there is no significant 31\,GHz emission originating at S1 (as seen in Figures\,\ref{fig:cbi-points} and \ref{fig:globalrgb}) implies that the radio-emission mechanism is enhanced by the local conditions in the PDR, otherwise, we would see higher levels of 31\,GHz emission near S1, given its large column of small grains (i.e. PAHs) \citep{cas08}.

The nature of the SD emissivity variations in $\rho$\,Oph may rely on various scenarios. SD models predict peak frequencies in the range 50-90\,GHz for the typical environmental conditions of a PDR \citep[e.g.][]{dra98, ysard+11, hensley+17}. However, only one reported observation of the California nebula is consistent with a SD spectrum with a (50 $\pm$ 17)\,GHz peak \citep{planckame14}. It is clear that the frequency of the SD spectrum must be changing due to environmental conditions across the PDR. In this case, the observed emissivity variations could be associated with a shift in the peak frequency. Nevertheless, the analysis in this work is based on the assumption that the 31\,GHz emission maps the AME.

A tentative explanation for the local emissivity boost, in the SD paradigm, could be that the main spin-up mechanism in $\rho$\,Oph\,W are ion collisions \citep{dl98b}. The electromagnetic coupling between passing C$^+$ ions and the grain dipole moment impart torques on the grain. As the rotational excitation by ions is more effective for charged grains \citep{hensley+17}, a possible explanation could be that the PDR is hosting preferentially charged PAHs spun-up by C$^+$. As observed in Fig.\,\ref{fig:rmap}, the fainter signal from S1 and SR3 in the radio maps of $\rho$\,Oph could then be naturally explained by the fact that these stars are too cold to create conspicuous C{\sc ii} regions \citep{cas08}. In addition to the rotational excitation of charged grains, we could also add the effect of a possible changing penetration of the ISRF through the PDR due to the morphology of the cloud. The variations in the ISRF intensity affect the grain size distribution and, most likely, their dipole moment, and these factors can produce considerable changes in the SD emissivity \citep{hoang11dipole}. Thus, further analysis of the PDR spectrum at different radio-frequencies is required to calculate its physical parameters and understand the causes of the detected SD emissivity variations.

\section{Conclusions}
\label{sec:conc}

We report 31\,GHz observations from CBI\,2 of the $\rho$\,Oph\,W PDR, at an angular resolution of 4.5\,arcmin. The emission runs along the PDR exposed to the ionizing star HD\,147889. Interestingly, there is no significant 31\,GHz emission from S1, the brightest IR nebula in the complex.

To understand the nature of the 31\,GHz emission, we calculated the correlations of the CBI\,2 data with different IR templates, with PAHs column density proxies, and with dust radiance and optical depth templates. We show that the 31\,GHz emission is related to the local PAH column density. The best correlation was found when using the 12\,$\upmu$m PAH column density proxy and it is significantly better than when using the 8\,$\upmu$m and 3\,$\upmu$m PAH column density proxies. 

We also measured the correlations at different angular resolutions and found that the dust radiance correlation is better at lower angular resolution. This shows that the effect of angular resolution must be considered when interpreting morphological correlations, implying in this case that PAH's cannot be ruled out as spinning dust carriers, as previous studies at lower angular resolutions have suggested. Additionally, we calculated a spectral index of $\alpha$\,=\,0.05\,$\pm$\,0.08, between 28.5\,GHz and 33.5\,GHz, that is consistent with the flat spectrum previously reported on this region. By using a 12\,$\upmu$m / 3\,$\upmu$m ratio map, we measured an increase of the PAH size in the transition between the PDR and the molecular cloud.

Motivated by the intrinsic scatter in the correlation plots, we constructed a proxy for the 31\,GHz emissivity per spinning dust emitter so as to quantify its variations. Using a {\tt gpu-uvmem} model of the 31\,GHz emission, we found that the spinning dust emissivity peak over the PDR is, at least, 25 times greater over the noise, at 3$\sigma$, and 23 times greater than at the location of star S1, also at 3$\sigma$. 

The spinning dust emissivity boost in $\rho$\,Oph\,W appears to be dominated by local conditions in the PDR. Based on the fainter 31\,GHz signal from stars S1 and SR3, possible explanations are environmental ions or a changing grain population. In the framework of the spinning dust interpretation, these emissivity variations may hold the key to identify the dominant grain spin-up mechanisms. Further multi-frequency radio analysis of the PDR spectrum is needed in order to better understand the conditions that give rise to spinning dust.

\section*{Acknowledgments} 
We thank the anonymous referee for helpful comments. CAT and SC acknowledge support from FONDECYT grant 1171624. MV acknowledges support from FONDECYT through grant 11191205. This work used the Brelka cluster (FONDEQUIP project EQM140101) hosted at DAS/U. de Chile. MC acknowledges support granted by CONICYT PFCHA/DOCTORADO BECAS CHILE/2018 - 72190574. CD was supported by an ERC Starting (Consolidator) Grant (No.~307209) under the FP7 and an STFC Consolidated Grant (ST/P000649/1). This work was supported by the Strategic Alliance for the Implementation of New Technologies (SAINT, see \url{www.astro.caltech.edu/chajnantor/saint/index.html}) and we are most grateful to the SAINT partners for their strong support. We gratefully acknowledge support from  B. Rawn and S. Rawn Jr. The CBI was supported by NSF grants 9802989, 0098734 and 0206416. This research has made use of data from the Herschel Gould Belt survey (HGBS) project (http://gouldbelt-herschel.cea.fr). 


\label{lastpage}

\bibliographystyle{mn2e}

\bibliography{roph_cbi2}
\bsp

\end{document}